\documentclass[fleqn,usenatbib]{rasti}

\usepackage{newtxtext,newtxmath}
\usepackage[T1]{fontenc}
\usepackage{array}
\DeclareRobustCommand{\VAN}[3]{#2}
\let\VANthebibliography\thebibliography
\def\thebibliography{\DeclareRobustCommand{\VAN}[3]{##3}\VANthebibliography}
\usepackage{natbib}

\usepackage{graphicx}
\usepackage{amsmath}
\usepackage{float}
\graphicspath{{images/}}
\usepackage{xcolor, soul}

\title[Measuring SN host-galaxy properties]{Using 4MOST to refine the measurement of galaxy properties: A case study of Supernova hosts}

\author[J. Dumayne et al.]{
J. Dumayne$^{1}$,
I. M. Hook$^{1}$,
S. C. Williams$^{2, 3}$,
G. A. Lowes$^{1, 4, 5}$,
D. Head$^{1}$,
A. Fritz$^{6}$,
O. Graur$^{7, 8}$,
\newauthor
B. Holwerda$^{9}$,
A. Humphrey$^{10, 11}$,
A. Milligan$^{1}$,
M. Nicholl$^{12}$,
B. F. Roukema$^{13, 14}$,
P. Wiseman$^{15}$.
\\
$^{1}$Physics department, Lancaster University, Bailrigg, Lancaster, LA1 4YB, UK\\
$^{2}$ Finnish Centre for Astronomy with ESO (FINCA), Quantum, Vesilinnantie 5, Univeristy of Turku, FI-20014 Turku, Finland\\
$^{3}$ Department of Physics and Astronomy, University of Turku, FI-20014 Turku, Finland \\
$^{4}$EA Milne Centre for Astrophysics, Department of Physics and Mathematics, University of Hull, HU6 7RX\\
$^{5}$Centre of Excellence for Data Science, Artificial Intelligence and Modelling, University of Hull, HU6 7RX\\
$^{6}$ OmegaLambdaTec GmbH, Lichtenbergstraße 8, 85748 Garching, Germany\\
$^{7}$ Institute of Cosmology and Gravitation, University of Portsmouth, Portsmouth PO1 3FX, UK\\
$^{8}$ Department of Astrophysics, American Museum of Natural History, Central Park West and 79th Street, New York, NY 10024, USA\\
$^{9}$ University of Louisville, Department of Physics and Astronomy, 102 Natural Science Building, 40292 KY Louisville, USA. \\
$^{10}$ DTx -- Digital Transformation CoLab, Building 1, Azur\'em Campus, University of Minho, 4800-058 Guimar\~aes, Portugal \\
$^{11}$ Instituto de Astrof\'isica e Ci\^encias do Espa\c{c}o, Universidade do Porto, CAUP, Rua das Estrelas, PT4150-762 Porto, Portugal \\
$^{12}$ Astrophysics Research Centre, School of Mathematics and Physics, Queens University Belfast, Belfast BT7 1NN, UK \\
$^{13}$ Institute of Astronomy, Faculty of Physics, Astronomy and Informatics, Nicolaus Copernicus University, Grudziadzka 5, 87-100 Toru\'n, Poland \\
$^{14}$ Univ Lyon, Ens de Lyon, Univ Lyon1, CNRS, Centre de Recherche Astrophysique de Lyon UMR5574, F--69007, Lyon, France\\
$^{15}$School of Physics and Astronomy, University of Southampton, Southampton SO17 1BJ, UK\\
}

\date{Accepted 03/08/2023. Received 28/07/2023; in original form 14/03/2023}

\pubyear{2023}

\begin{document}
\label{firstpage}
\pagerange{\pageref{firstpage}--\pageref{lastpage}}
\maketitle

\begin{abstract}
The Rubin Observatory's 10-year Legacy Survey of Space and Time will observe near to 20 billion galaxies. For each galaxy the properties can be inferred. Approximately $10^5$ galaxies observed per year will contain Type Ia supernovae (SNe), allowing SN host-galaxy properties to be calculated on a large scale. Measuring the properties of SN host-galaxies serves two main purposes. The first is that there are known correlations between host-galaxy type and supernova type, which can be used to aid in the classification of SNe. Secondly, Type Ia SNe exhibit correlations between host-galaxy properties and the peak luminosities of the SNe, which has implications for their use as standardisable candles in cosmology. We have used simulations to quantify the improvement in host-galaxy stellar mass ($M_\ast$) measurements when supplementing photometry from Rubin with spectroscopy from the 4-metre Multi-Object Spectroscopic Telescope (4MOST) instrument. We provide results in the form of expected uncertainties in $M_\ast$ for galaxies with 0.1 < $z$ < 0.9 and 18 < $r_{AB}$ < 25. We show that for galaxies mag 22 and brighter, combining Rubin and 4MOST data reduces the uncertainty measurements of galaxy $M_\ast$ by more than a factor of 2 compared with Rubin data alone. This applies for elliptical and Sc type hosts. We demonstrate that the reduced uncertainties in $M_\ast$ lead to an improvement  of 7\% in the precision of the "mass step" correction. We expect our improved measurements of host-galaxy properties to aid in the photometric classification of SNe observed by Rubin. 

\end{abstract}

\begin{keywords}
instrumentation -- transients: supernovae  -- techniques: spectroscopic
\end{keywords}


\section{Introduction}
The 10 year Legacy Survey of Space and Time (LSST) \citep{09LSST} conducted by the Vera Rubin Observatory will observe 20 billion galaxies over the 10-year survey\footnote{https://www.lsst.org/scientists/keynumbers}. Similarly, the 4-metre Multi-Object Spectroscopic Telescope
(4MOST) will collect 13 million spectra of galaxies \footnote{https://www.4most.eu/cms/science/exgalconsurv/}. Amongst these observed galaxies will be a large quantity of galaxies which host transients. Both of these surveys will be carried out by next generation facilities (see sections \ref{Vera Rubin} and \ref{4most intro}). A previous large survey, the Sloan Digital Sky Survey, collected 1.5 million galaxy spectra (e.g. \citealt{York00, Alam15}). The next generation surveys will allow the calculation of galaxy properties on a larger scale than ever before.
\\\indent Since the early observational evidence that the Friedmannian scale factor of the Universe is accelerating and definitive evidence via SNe Ia  \citep{Reiss1998, Perlmutter1999}, there have been many attempts to understand the cause of the acceleration. Dark energy is often invoked as an explanation and recent work in cosmology has focused on measuring the dark energy equation of state parameter, $w$ (e.g., \citealt{Garnavich98, Scolnic18, 2022Chen}). A popular method uses standardised SNe Ia light curves, which allows SNe Ia to be used as distance indicators (e.g., \citealt{92Branch, 08Wood, 2022Dhawan}). The measured luminosity  distances and redshifts of a sample of SNe Ia can therefore be used to constrain the cosmological parameters, including $w$ \citep{04Riess, 07Riess, 08Kowalski, 2022Brout}. 
\\\indent For each SN observed by the Vera Rubin Observatory, it will be possible to calculate the properties of the host galaxy using Spectral Energy Distribution (SED) fitting to multi-colour photometric measurements (e.g., \citealt{Spinrad72, 13Conroy, Kelsey20}). Incorporating spectral information would facilitate a more accurate calculation of host-galaxy properties. \citet{Childress13} demonstrated the power of spectra observed by the Nearby Supernova Factory \citep{2002Aldering} combined with UV from Galex \citep{2007Morrissey}. However, for larger samples this will not be possible. An alternative to using SED fitting would be to use machine learning. The \citet{Humphrey22} demonstrated that  transfer learning is better at recovering galaxy properties than SED fitting, when only broadband photometry is used. In our work we consider spectra observed by 4MOST. 4MOST is ideally suited to spectroscopy of large samples of SN host-galaxies, due to its high multiplex (see Section \ref{4most intro}).
\\\indent There are several reasons for wanting to obtain more precise host-galaxy properties. The first is that host-galaxy properties correlate with the peak magnitude of SNe Ia (e.g., \citealt{Kelly10, Lampeitl2010, Sullivan2010}), this is often called the "mass step", and hence corrections need to be made in order to avoid biases in measurements of cosmological parameters. The second reason for wanting more precise host-galaxy properties is because there are correlations between SN classes and host-galaxy properties (e.g., \citealt{Hamuy01, Galbany14, Gagliano21}), which can be used to aid classification of SNe. Additionally, measuring host-galaxy properties with more precision will lead to a better constraint on the V-band extinction of a galaxy \citep{Tonry03, Holwerda08, Holwerda15}. Finally, improved measurements of host-galaxy properties will lead to a more accurate measurement of demographics of galaxies and the populations of transients within them. This will lead to a better understanding of the dependence of transient type on host-galaxy properties.
\\\indent The mass step is observed when the Hubble residual (the difference between the distance modulus to the SN and the predicted value by a cosmological model at the SN's redshift (e.g., \citealt{Jha07, Gallagher08, Kelly10}) is plotted as a function of the host-galaxy stellar mass ($M_\ast$). Observing the Hubble residual as a function of host-galaxy $M_\ast$ shows that SNe Ia in high mass galaxies (>$10^{10}M_\odot$) are brighter than SNe Ia in low mass galaxies (<$10^{10}M_\odot$), after correction for stretch and colour \citep{Childress13}. Empirical evidence has shown the mass step appears at approximately $10^{10}M_\odot$ \citep{Sullivan2010, Uddin17}. Therefore a correction can be applied as a step function, depending on which side of $10^{10}M_\odot$ the host-galaxy's $M_\ast$ falls.
\\\indent A SN host-galaxy's $M_\ast$ is not the only property that can lead to a correction being applied to a SN. \citet{Gallagher08} found a correlation between the Hubble residual and metal abundance. Additionally, \citet{Wolf16} and \citet{2020Rigault} found a correlation between Hubble residual and specific star formation rate. \citet{22Briday} and \citet{Wiseman23} demonstrate stellar population age is the galaxy parameter that drives the step. Metallicity, star-formation rate and host-galaxy mass are linked, so these results are to be expected \citep{2014Speagle, 2022Li}. \citet{Brout21} report that the mass step can be explained by introducing a new SN colour model, by modelling different dust distributions. Many of these properties can be measured by fitting a combination of photometric and spectroscopic data of the host galaxies \citep{2022Jones, 2020Lower}.
\\\indent Host-galaxy properties can also be used to aid the classification of transients, especially in cases where  spectroscopy of the transient is not available \citep{Foley13, Pan14}. SNe Ia appear more frequently in star-forming galaxies than passive galaxies (e.g., \citealt{Oemler79, Botticella17}), with 10 times as many SNe Ia appearing in strongly star-forming galaxies compared with passive galaxies \citep{Sullivan06}. The work of \citet{2017Graura} found that SNe Ia rates anticorrelate with the host-galaxy mass. Their follow up work confirmed that SNe Ia are more common in low-mass galaxies \citep{2017Graurb}. Type Ia, Ib/c and II SNe are more common in late-stage spiral galaxies than early-stage spiral galaxies \citep{Mannucci05}. Type Ic SNe have host-galaxies with high specific star formation rates and low metallicities \citep{Modjaz20}. Therefore by measuring the properties of host-galaxies such as star formation rate, we hope to be able to improve the classification of transients from LSST even when there is no transient spectrum available.

\subsection{The Vera C. Rubin Observatory} \label{Vera Rubin}
\indent The Vera C. Rubin observatory is expected to begin collecting observations for LSST in 2025. The observatory will survey a large portion of the southern hemisphere. The main telescope of the Rubin observatory is the Simonyi Survey Telescope. The telescope has an 8.4 metre primary mirror, with the world's largest CCD camera \footnote{The specifications for the Simonyi Survey Telescope can be found on the  Rubin observatory website (https://www.lsst.org/about).}. The Rubin Observatory will observe a 9.62 square-degree area of sky at a single pointing,\footnote{https://www.lsst.org/about/tel-site/optical\_design} with the entire survey covering $\sim$30,000\, deg$^2$ \citep{Ivezic2019}. The Wide, Fast, Deep survey, which is the primary survey, will have a declination range of -65 to +5 degrees \citep{LSST_obs}, although it should be noted that the final survey design has not been decided. The Rubin Observatory is expected to observe about $10^5$ SNe Ia per year for 10 years \citep{LSSTDoc}, of which a total amount of approximately 112,000 will be suitable for cosmology \citep{Mandelbaum18}. In this paper we are concerned with measurements of the host galaxies. In order to estimate the quality of photometry from LSST, we assume the 10-year, 5-$\sigma$ survey depths of the Wide Fast Deep survey, namely u=26.1, g=27.4, r=27.5, i=26.8, z=26.1 and y=24.9 mags.\footnote{https://www.lsst.org/scientists/keynumbers} We use AB magnitudes throughout this paper.

\subsection{4MOST} \label{4most intro}
\indent 4MOST is a new high-multiplex, wide-field spectroscopic survey facility under development for the 4m VISTA Telescope \citep{Guiglion19, deJong19a}. 
4MOST is due to begin operations in 2024. It will be a fibre-fed spectrograph, with 2,436 fibres in an approximately 4 square degree field-of-view.\footnote{The specifications for 4MOST can be found on the ESO website (www.eso.org/sci/facilities/develop/instruments/4MOST.html\#BasSpec).}. Each fibre will have a diameter of 1.45 arcseconds \footnote{Details about 4MOST's fibres can be found in the 4MOST user manual https://www.4most.eu/cms/facility/overview/.}. A third of the fibres will be connected to a High-Resolution Spectrograph (HRS), with the remaining two-thirds of fibres being connected to two Low-Resolution Spectrographs (LRS).  The LRS will observe with a resolution of 5,000 at a wavelength range of 370-950 nm. The HRS will have a resolution of approximately 20,000 and will observe at 392.6-435.5, 516-573 and 610-679 nanometres \citep{2016DeJong}. In a 5-year survey 4MOST will be able to cover approximately 21,000 square-degrees of sky, covering a declination range of -70 < dec < 5 degrees \citep{4MOST_obs}. This declination range has a significant overlap with LSST. During this survey approximately 20 million low-resolution and 3 million high-resolution spectra will be observed.\footnote{See footnote 6.} This assumes an exposure time of 2 hrs. The details of the 4MOST survey are still to be decided.
\\\indent The Time Domain Extragalactic Survey (TiDES) will carry out spectroscopic follow up of photometrically observed transients. TiDES aims to collect 35,000 live transients and 50,000 host galaxy observations during the first 5 years of 4MOST (\citealp{Tides2019}; Frohmaier et al. (in prep)). This approach has been used successfully by the Dark Energy Survey (DES). DES supplemented deep host-galaxy photometry \citep{Wiseman20} with fibre-fed spectroscopy from the Anglo-Australian Telescope \citep{Lidman20}.

\subsection{Aims of this work}
\indent This research aims to investigate the extent to which the precision of measuring a galaxy's $M_\ast$ can be improved using 4MOST with photometry, compared with using photometry alone. This paper is organised as follows. Section \ref{Method} presents our method of producing a target spectrum, synthetic photometry and synthetic 4MOST spectra. Section \ref{Results} shows the results obtained by this research, and Section \ref{Discussion} analyses the results. Finally, we conclude in Section \ref{Summary}.

\section{Method} \label{Method}
As 4MOST and the Vera Rubin Observatory are not operational yet, we use simulated data. To assess both instruments' ability to measure galaxy properties, we need to start with a spectrum of a galaxy with values of physical properties that we adopt as the ground truth for this experiment. We use an example output spectrum of the Fitting Assessment of Synthetic Templates (FAST) SED fitting code \citep{Kriek2009}, which has associated $M_\ast$, star formation rate and other galaxy property values. In order to generate the target galaxy properties with which we can compare our results, we ran FAST for an initial pass (we later use FAST in a second pass to analyse the simulated 4MOST output). For this we gave FAST an input spectrum to fit. We used the \citet{96Kinney} elliptical template which is built into the 4MOST Exposure Time Calculator (4MOST ETC)\footnote{We used the internal \textsc{python} based 4MOST ETC; nonetheless, the public web-based ETC provides the same results (\href{https://etc.eso.org/observing/etc/fourmost}{https://etc.eso.org/observing/etc/fourmost}).}. In later stages of the work we carry out the process with an Sc galaxy, however the initial results are obtained for the elliptical galaxy. We only used an elliptical and an Sc galaxy. As elliptical galaxies have the least amount of star formation \citep{2017Kokusho}, comparing this to a star forming spiral galaxy will allow us to test the most extreme situations. In the following Section we describe the workflow from the \citeauthor{96Kinney} spectrum to the template spectrum with known properties. This also provides an overview of the method which will be elaborated in the remainder of Section \ref{Method}.

\subsection{Producing a template spectrum}
\label{making template}
We begin with the \citet{96Kinney} elliptical spectrum redshifted to 0.3, approximately the middle of the expected TiDES redshift range. The \citeauthor{96Kinney} spectrum was used as it had the best signal-to-noise ratio and was readily available. The elliptical galaxy spectrum was normalised to r = 21 and an observation was simulated using the 4MOST ETC code (see Section \ref{Making 4MOST} for the parameters assumed). The raw output from the 4MOST ETC and the corresponding noise spectrum were processed to generate realistic-flux-calibrated spectra (see Section \ref{Making 4MOST} for details). Synthetic photometry was also created, and the process to do this is explained in Section \ref{Making phot}.
\\\indent The spectrum and photometry were then fit with FAST. FAST takes a parameter file which defines the settings to be applied, as shown in table \ref{FAST table}. Once defined these are not changed. This initial run through FAST produced a best-fit spectrum that acts as the template which later fitted-galaxy-property values are compared with. From this calculated value we could extrapolate additional target values for redshift = 0.3, by using the proportional relationship of log($M_\ast$) and magnitude.

\begin{table}
\begin{center}
\begin{tabular}{ |c|c| } 
 \hline
 Parameter & Option or Range chosen \\ 
 \hline \hline
 Number of simulations & 10,000 \\
 \hline
 Confidence interval & 68\% \\ 
 \hline
 Stellar population library & \citet{03Bruzual}\\
 \hline
 Stellar initial mass function & \citet{03Chabrier} \\ 
 \hline
 Star formation history & Delayed exponential SFH \\
 \hline
 SFR Average & 0 (Instantaneous SFR) \\
 \hline
 Method to find best-fit & Median of Monte Carlo \\
 \hline
 Dust law & \citet{13Kriek} \\
 \hline
 log($\tau$) [log($\tau$/yr)] & 6.5 - 11 \\ 
 \hline
 log(age) [log(age/yr)] & 8.0 - 9.8 \\
 \hline
 V-band extinction (A$_V$) & 0.0 - 3.0 \\ 
 \hline
 Metallicity & 0.004, 0.008, 0.02, 0.05 \\
 \hline
 Hubble Constant [km/s/Mpc] & 70.0 \\ 
 \hline
 $\Omega_{M}$ & 0.3 \\
 \hline
 $\Omega_{\Lambda}$ & 0.7 \\
 \hline
\end{tabular}
\caption{The parameters that we chose to use when running FAST to find the best-fitting galaxy parameters for input photometry and/or spectra.}
\label{FAST table}
\end{center}
\end{table}

\subsection{FAST} \label{FAST}
FAST works by fitting stellar population synthesis templates to a spectrum and/or broadband photometry. FAST then returns values for the galaxy properties of the best-fit galaxy. The galaxy properties FAST can calculate are: redshift, metallicity, stellar age, V-band extinction (A$_V$), $M_\ast$, star formation rate, specific star formation rate, star formation timescale ($\tau$) and the ratio of age to star formation timescale. As mentioned previously, FAST takes a parameter file that defines the allowed ranges of each parameter over which it searches (relating to FAST's library of galaxy spectra). The chosen parameters can be seen in Table \ref{FAST table}. The best-fitting galaxy properties are found by taking the median of the distribution of 10,000 runs of FAST. In each run FAST alters the photometry and/or spectral flux values within the corresponding error values. FAST then fits these new photometry values in each run. After 10,000 runs, there is a range of calculated galaxy property values. FAST then finds the upper and lower limits that contain 68$\%$ of the data, to find the equivalent (Gaussian interpretation) 1-$\sigma$ uncertainty range for each galaxy property. FAST does not use a minimum-searching algorithm, instead it fits every point of the parameter space.\footnote{https://github.com/jamesaird/FAST} FAST was used for this project due to its ability to fit both a spectrum and photometry.

\subsection{Synthetic photometry} \label{Making phot}
\indent The next stage of the simulations takes the noiseless-template spectrum (from Section \ref{making template}) to produce synthetic photometry and a spectrum as observed by 4MOST. During the work described in this Section and \ref{Making 4MOST}, we shifted the template to different redshifts (0.1 < $z$ < 0.9) and normalised it to different desired magnitudes in the range (18 < r < 25). We call these shifted target spectra. For each case the \textsc{Python} package \textsc{Pyphot} (version 1.0) was used to calculate the flux within the filter bands u, g, r, i, z and y. The throughput received after passing through the atmosphere, the filters and the detectors can be seen in Figure \ref{fig: mag_flux}.  \textsc{Pyphot} calculates this by integrating each template through the given filter bands. The Rubin filters provided on the Vera Rubin Observatory's website \footnote{https://www.lsst.org/scientists/keynumbers} are used for this process, which we have added into both \textsc{Pyphot} and FAST. For each filter band we calculate the 1-$\sigma$ sky noise corresponding to the 10-year LSST depth, by scaling from the 5-$\sigma$ depths given in Section \ref{Vera Rubin}, assuming that sky noise dominates the photometric error. For each photometry point, we use the sky noise in the corresponding filter as the photometric uncertainty, unless the sky noise is less than 1\% of the flux, in which case we set the photometric error to be 1\% of the flux. This ensures that we do not use unrealistically small photometry errors.

\subsection{Simulated 4MOST spectra} \label{Making 4MOST}
\begin{figure}
    \centering
    \includegraphics[scale=.22]{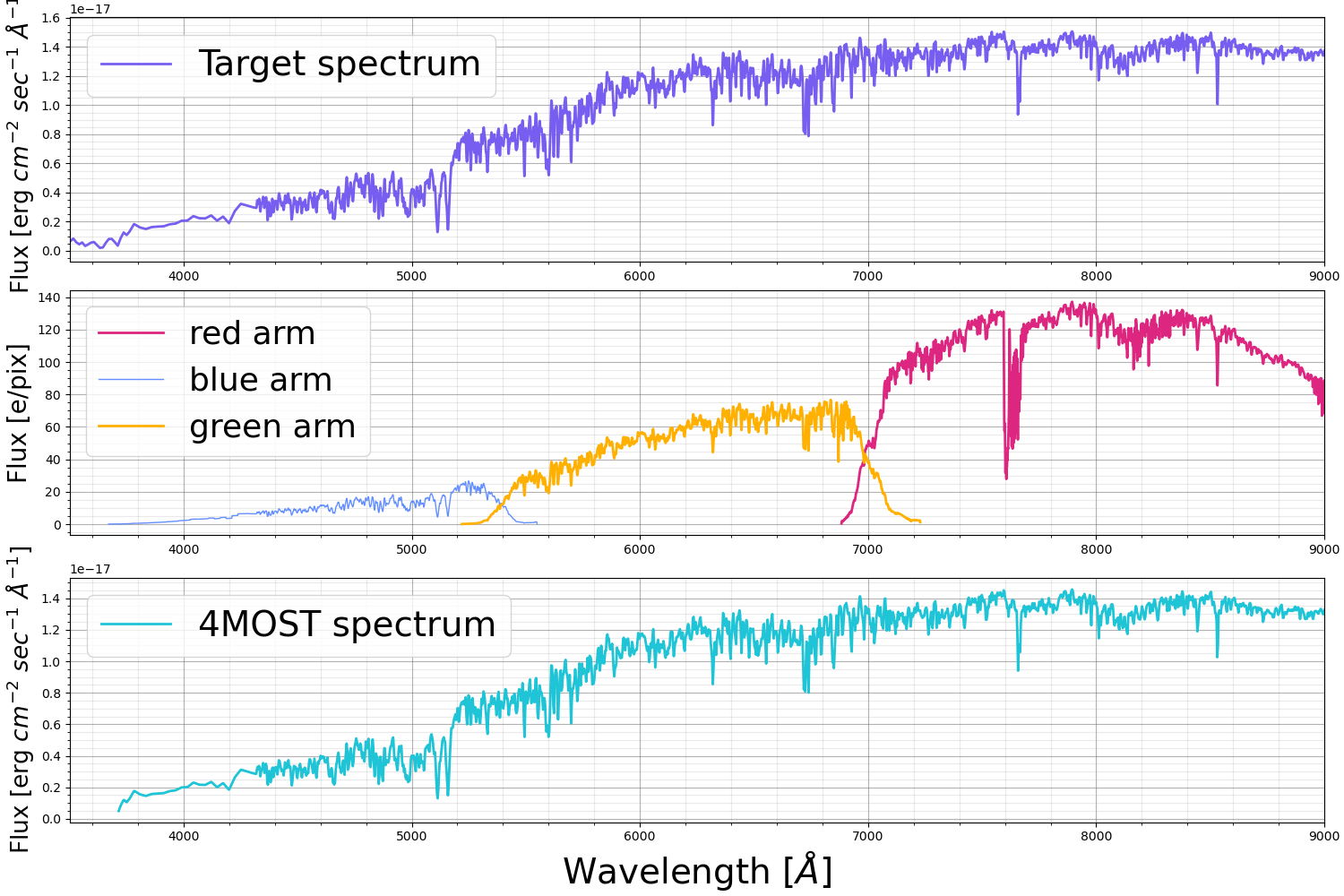}
    \caption{The spectrum at different stages throughout the process of simulating a spectrum from 4MOST. The top plot is the best-fit spectrum produced during the target stage. The middle plot is the target best-fit spectrum after it has been put through the 4FS ETC to produce a noiseless spectrum. Finally, the bottom plot is the flux-calibrated spectrum. This is the flux-calibrated object spectrum after it has been binned and combined into a single spectrum.}
    \label{fig: 4most stages}
\end{figure}

\indent The template spectrum is redshifted and magnitude normalised to each of the required values, then input into the 4FS ETC. The 4FS ETC is a software tool that estimates what 4MOST would see with specific observing conditions given an input spectrum with a specific magnitude. We use version 2.04. of the 4FS ETC. There is a newer version of the 4FS ETC which was released during the process of this work, version 2.2. The newer version produces additional outputs compared to version 2.04. (which we do not use). The outputs used from version 2.04 were compared to the  equivalent output from version 2.2. and found to be identical. The observing conditions were kept constant during this study. The 4FS ETC was set to have an airmass of 1.2, seeing of 0.8 arcseconds and a dark moon. We use an exposure time of 2 hrs. Later in the process we allow for the fact that brighter objects will be removed from the observing queue after reaching the spectral success criterion (see the final paragraph in this Section).
\\\indent The 4FS ETC v2.04. produces: spectrograph gain [electrons/adu], target signal count [electrons], sky background count [electrons], CCD dark current [electrons], CCD readout noise [electrons], noise count [electrons], efficiency [electrons/photon] and spectral bin width [nm]. All of these are produced as a spectrum with wavelength units of nanometres, which we convert to the units of angstrom. We will call the target-signal-count spectrum 'object spectrum', to prevent confusion with our shifted target spectrum. The object spectra are produced separately for blue, green and red wavelength ranges corresponding to the three arms of the 4MOST spectrograph. The top and middle panel of Figure \ref{fig: 4most stages} show examples of input and output of the 4FS ETC.

\begin{figure}
    \centering
    \includegraphics[scale=.32]{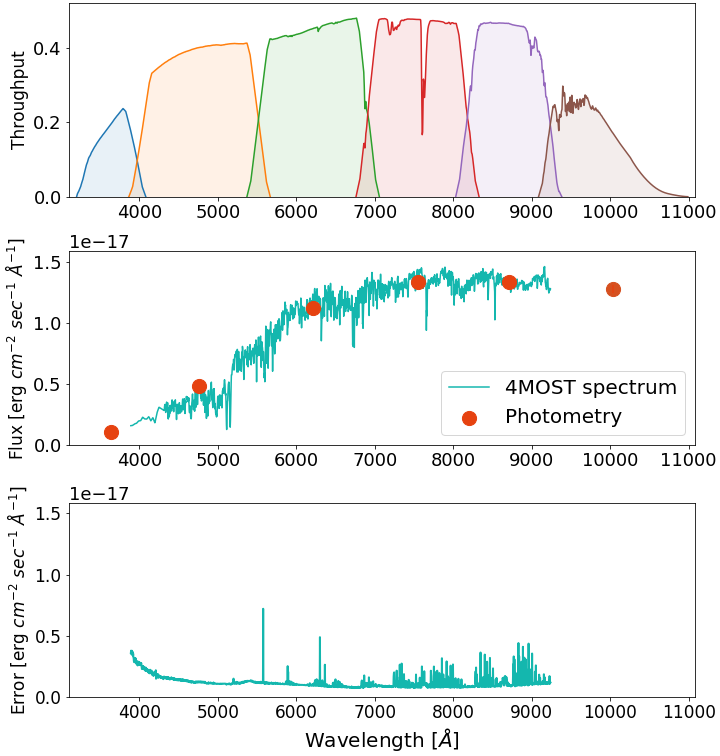}
    \caption{The top panel shows the throughput receieved, after passing through the atmosphere, filters and detectors. The middle panel shows an example of the 4MOST spectra flux produced by the 4MOST ETC and combined into a single spectrum for magnitude 21, $z$ = 0.3. The photometry for magnitude 21 can also be seen as the circles plotted on top of the 4MOST spectrum. The photometry uncertainties are also plotted, however they are smaller than the photometry symbol and so cannot be seen. Finally, the lower plot is the error values that correspond to the spectrum.}
    \label{fig: mag_flux}
\end{figure}

\indent To mimic flux calibration we multiplied each object spectrum by the corresponding gain, and then divided by the instrument response function. In the absence of simulated-spectral-standard stars we determined the response function using the shifted target spectrum itself, as shown as follows
\begin{eqnarray}
\mathrm{res}(\lambda_i) = \frac{\mathrm{obj}(\lambda_i). g(\lambda_i)}{\mathrm{targ}(\lambda_i) . t_{\mathrm{exp}}}
\end{eqnarray}

\noindent where res$(\lambda_i)$ is the response at the wavelength of the $i$th pixel, obj$(\lambda_i)$ is the object spectrum, $t_{\mathrm{exp}}$ is exposure time, $g$ is gain at each wavelength value, targ$(\lambda_i)$ is the shifted target spectrum before it was input into the ETC. This effectively assumes perfect flux calibration. In practice we expect 4MOST to produce relative flux calibration (which is what is important to this appliction), and absolute calibration will be provided by comparison to LSST photometry. The object spectra produced by the ETC are noiseless (however the ETC also produces noise values for each wavelength value). A calibrated noise spectrum is calculated for each of the three wavelength ranges by dividing the noise values for each wavelength by its corresponding response value.

\indent To calculate the galaxy's properties using FAST, the three sections of each object spectrum and calibrated noise spectrum must be combined into a continuous spectrum. Since the three arms of 4MOST have different wavelength binning, a binning function is used to place all three sections on a regular wavelength step of 3 \AA. This binning process is also applied to the weight spectra. The weight spectrum, w$(\lambda_j)$, is defined as follows
\begin{eqnarray}
w(\lambda_j) = \left( \frac{n(\lambda_j)}{\mathrm{res}(\lambda_j)} \right)^{-2}
\end{eqnarray}

\noindent where $n(\lambda_j)$ is the noise spectrum and all spectra are defined on the new wavelength spacing, $\lambda_j$. The weighting function ensures the extreme start and end values of a spectrum are given less weighting, where the throughput of the instrument is small. The weighting function was then applied to combine the spectra in the regions, where the spectra overlap through a weighted average, i.e.
\begin{eqnarray}
f(\lambda_j) = \frac{w_1(\lambda_j) . \mathrm{obj}_1(\lambda_j) . g_1(\lambda_j) + w_2(\lambda_j) . \mathrm{obj}_2(\lambda_j) . g_2(\lambda_j)} {w_1(\lambda_j) + w_2(\lambda_j)}
\end{eqnarray}

\begin{figure} 
    \centering
    \includegraphics[scale=.39]{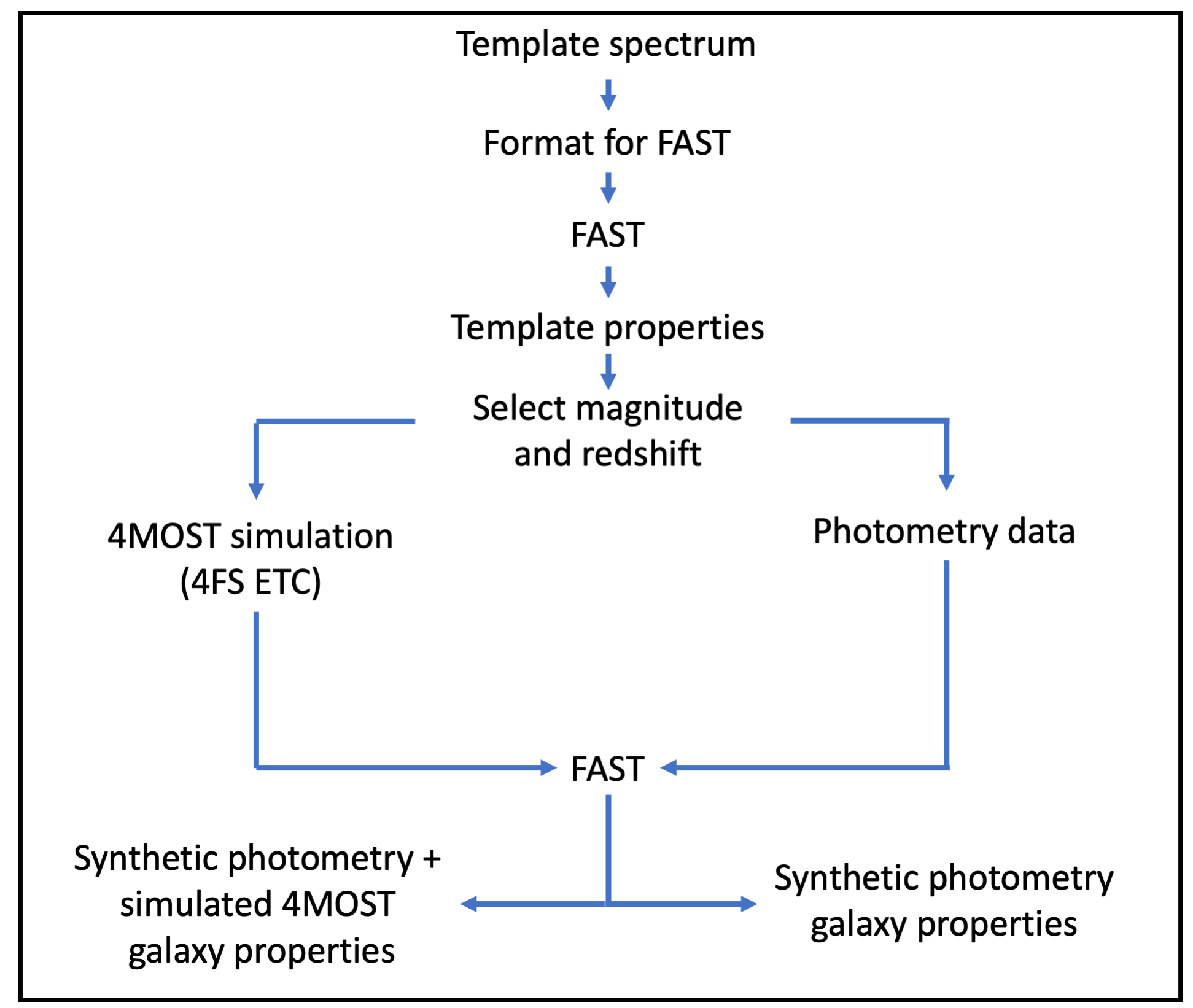}
    \caption{The steps carried out to calculate galaxy properties for magnitudes 18 < r < 25 and redshifts 0.1 < $z$ < 0.7.}
    \label{fig: flowchart}
\end{figure}

\begin{figure}
    \centering
    \includegraphics[scale=.46]{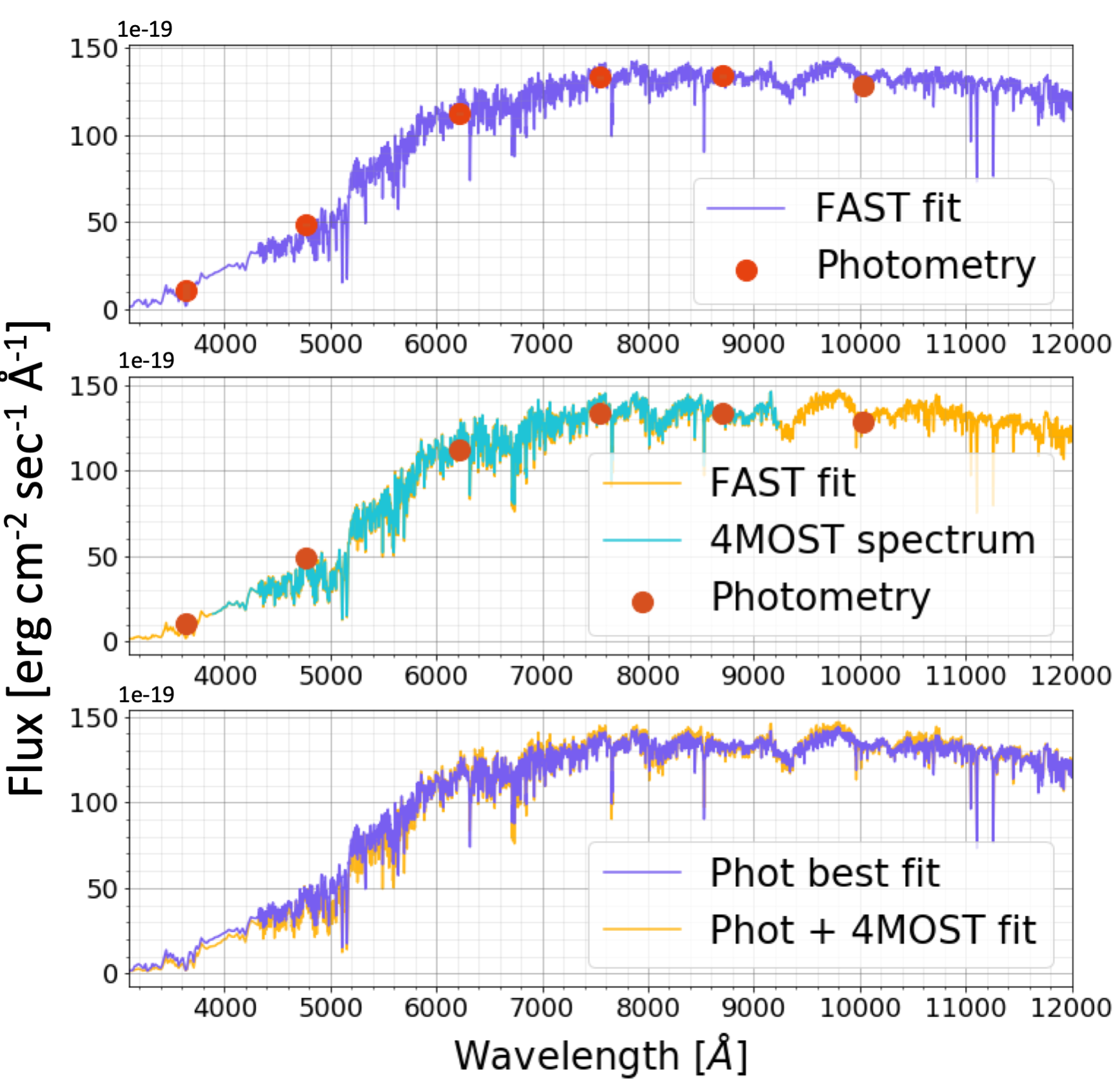}
    \caption{The top panel shows the best-fit found by FAST for photometry alone (purple line), with the photometry points plotted (orange circles). Full profiles of the Rubin filters are shown in Figure \ref{fig: mag_flux}. The middle panel shows the best-fit found for photometry with a 4MOST spectrum (gold line). The photometry is plotted on top (orange circle) with the 4MOST spectrum (blue line). The bottom panel shows a comparison of the two best-fits. As can be seen the two best-fits diverge below 6000 angstroms. All of the plots are for an elliptical galaxy at magnitude 21 and redshift 0.3.}
    \label{fig: FAST best-fit}
\end{figure}

\noindent where $f(\lambda_j)$ is the final combined flux, obj$_1$ and obj$_2$ are the two spectra to be joined, $w_1$ and $w_2$ are their corresponding weight spectra and $g_1$ and $g_2$ are their corresponding gain spectra. The error values in the overlap section are summed in quadrature to produce a single continuous error spectrum. An example of the final combined spectrum is shown in the bottom panel of Figure \ref{fig: 4most stages}.

\indent Once the photometry and 4MOST spectrum have been produced, along with their corresponding values, they are input into FAST for the main comparison between photometry only (referred to from here on as ``phot'') and photometry with 4MOST spectra (referred to from here on as ``phot + 4MOST''). An example of the input is shown in Figure \ref{fig: mag_flux}. One version of results was produced from inputting the photometry in to FAST alone and another version of results was produced by inputting the photometry and 4MOST spectrum together. A flowchart of the process can be seen in Figure \ref{fig: flowchart}. The photometry and spectrum input into FAST, compared with the best-fit produced by FAST can be seen in Figure \ref{fig: FAST best-fit}. To account for bright objects being removed from the queue we assumed an uncertainty floor. An uncertainty cannot be smaller than what has been observed when a galaxy reaches a signal-to-noise of 3. When an object reaches this value it is removed from 4MOST's observing queue. The magnitude for each galaxy type to reach the signal-to-noise criteria, for each redshift, was measured. For each galaxy property calculated from spectra, the recorded uncertainty at this magnitude was applied to all brighter objects of the same redshift.

\begin{figure*}
    \centering
    \includegraphics[scale=.38]{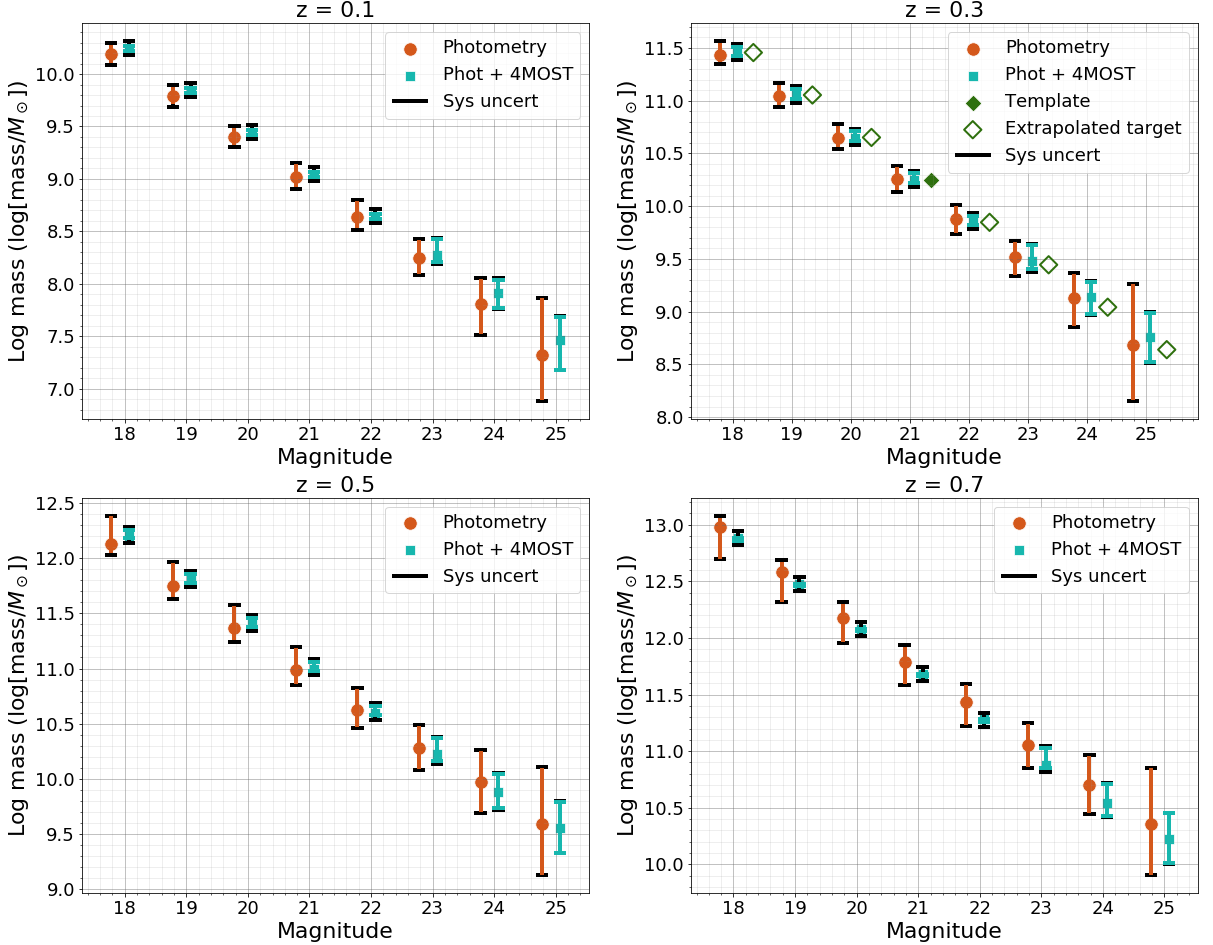}
    \caption{Derived log($M_\ast$) as a function of magnitude for LSST photometry only (orange circles) and for LSST phot + 4MOST spectroscopy combined (blue squares) for an elliptical galaxy at $z$=0.3. The template spectrum's $M_\ast$ is also shown (filled green diamond). The template mass at other magnitudes were extrapolated (empty green diamonds) from the original template spectrum (filled green diamond). There is an uncertainty associated with the template spectrum, which is the uncertainty that FAST calculates from the input real galaxy spectrum, although the uncertainty is too small to see in this Figure. The extended uncertainties (black line) show the total error bar when an estimate of systematic uncertainty is added in quadrature. The log ($M_\ast$) of the galaxies are shown as a function of magnitude and redshift. It can be seen that at all simulated redshifts and magnitudes the precision of the calculated galaxy mass is significantly improved for phot + 4MOST, compared with using photometry alone.}
    \label{fig: 4panel mass}
\end{figure*}

\begin{figure*}
    \centering
    \includegraphics[scale=.38]{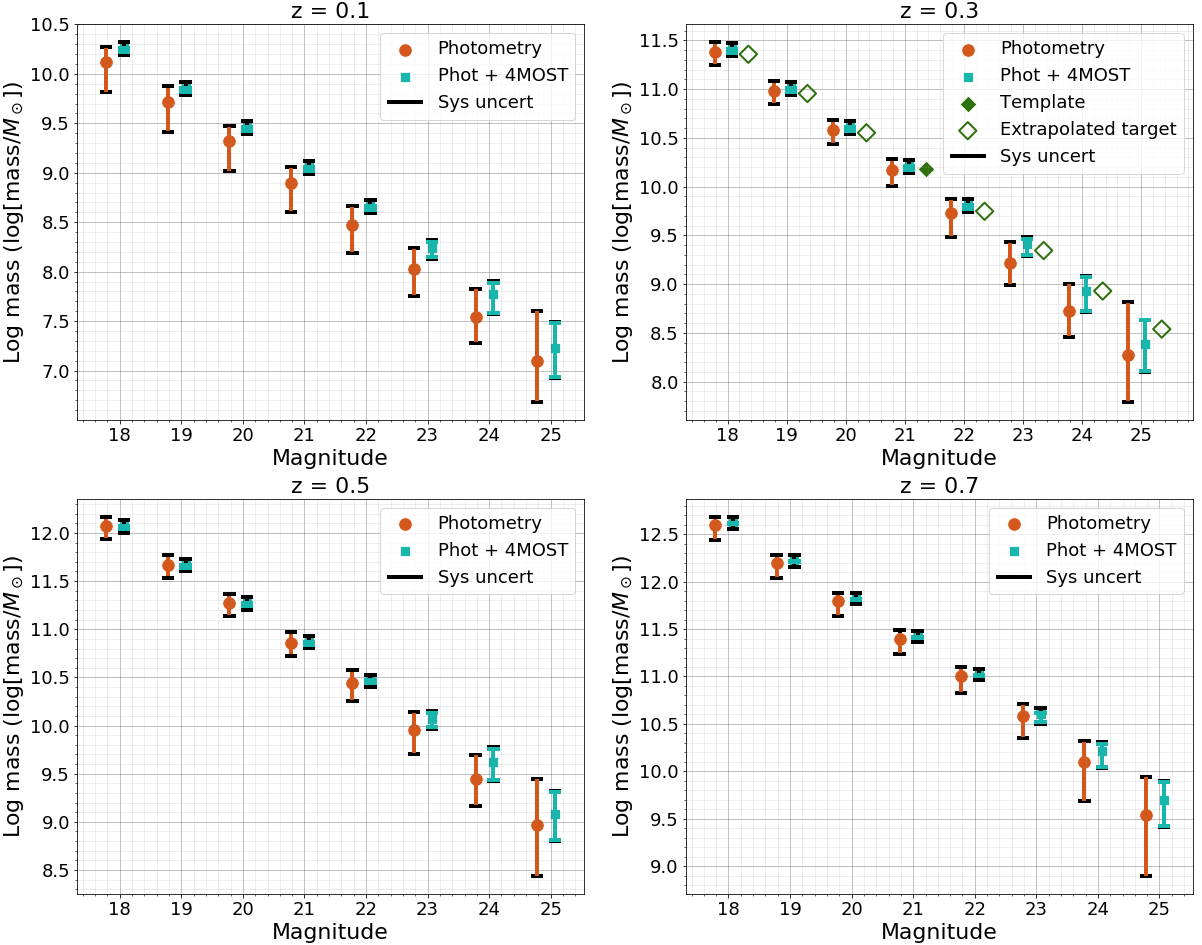}
    \caption{The same as figure \ref{fig: 4panel mass} but for a Sc galaxy. It can be seen that at all simulated redshifts and magnitudes the precision of the calculated galaxy mass is significantly improved for phot + 4MOST, compared with using photometry alone.}
    \label{fig: 4panel mass sc}
\end{figure*}

\section{Results} \label{Results}
\begin{figure}
    \centering
    \includegraphics[scale=.35]{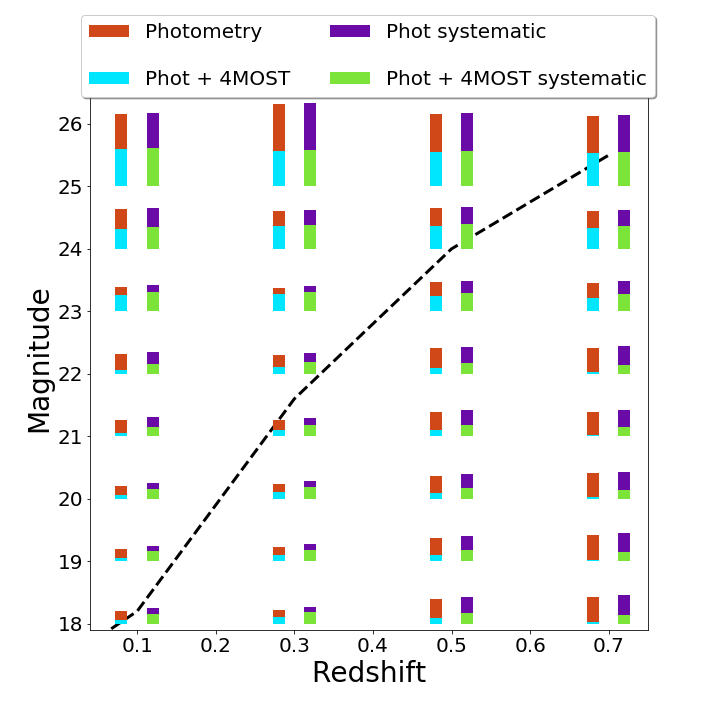}
    \caption{A comparison of the uncertainty on log(mass/$M_{\odot}$) for a an elliptical galaxy for a collection of redshifts and magnitudes when phot + 4MOST are used together can be seen in light blue. The additional uncertainty measured when using photometry only can be seen in orange. The size of the line is proportional to the size of the uncertainty. The dashed black line shows where the $M_\ast$ of the galaxy is $10^{10}$ $M_{\odot}$. Finally, the systematic uncertainty from using SED fitting is added in quadrature to be compared with the uncertainty values. The phot + 4MOST with systematic uncertainties is shown in green. The additional uncertainty for phot with systematic uncertainties is shown in purple.}
    \label{fig: circle el}
\end{figure}

\begin{figure}
    \centering
    \includegraphics[scale=.35]{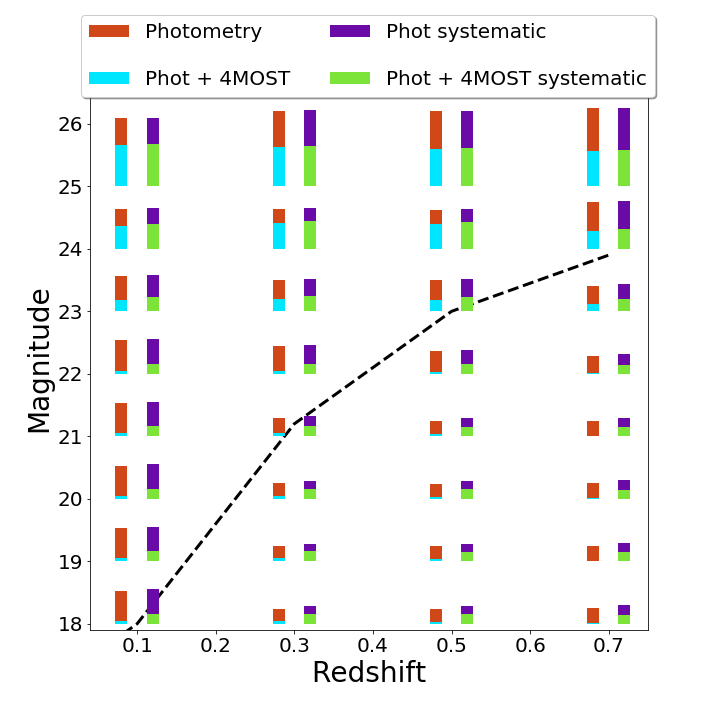}
    \caption{The same as Figure \ref{fig: circle el} but for a Sc galaxy.}
    \label{fig: circle sc}
\end{figure}

\begin{figure*}
    \centering
    \includegraphics[scale=.38]{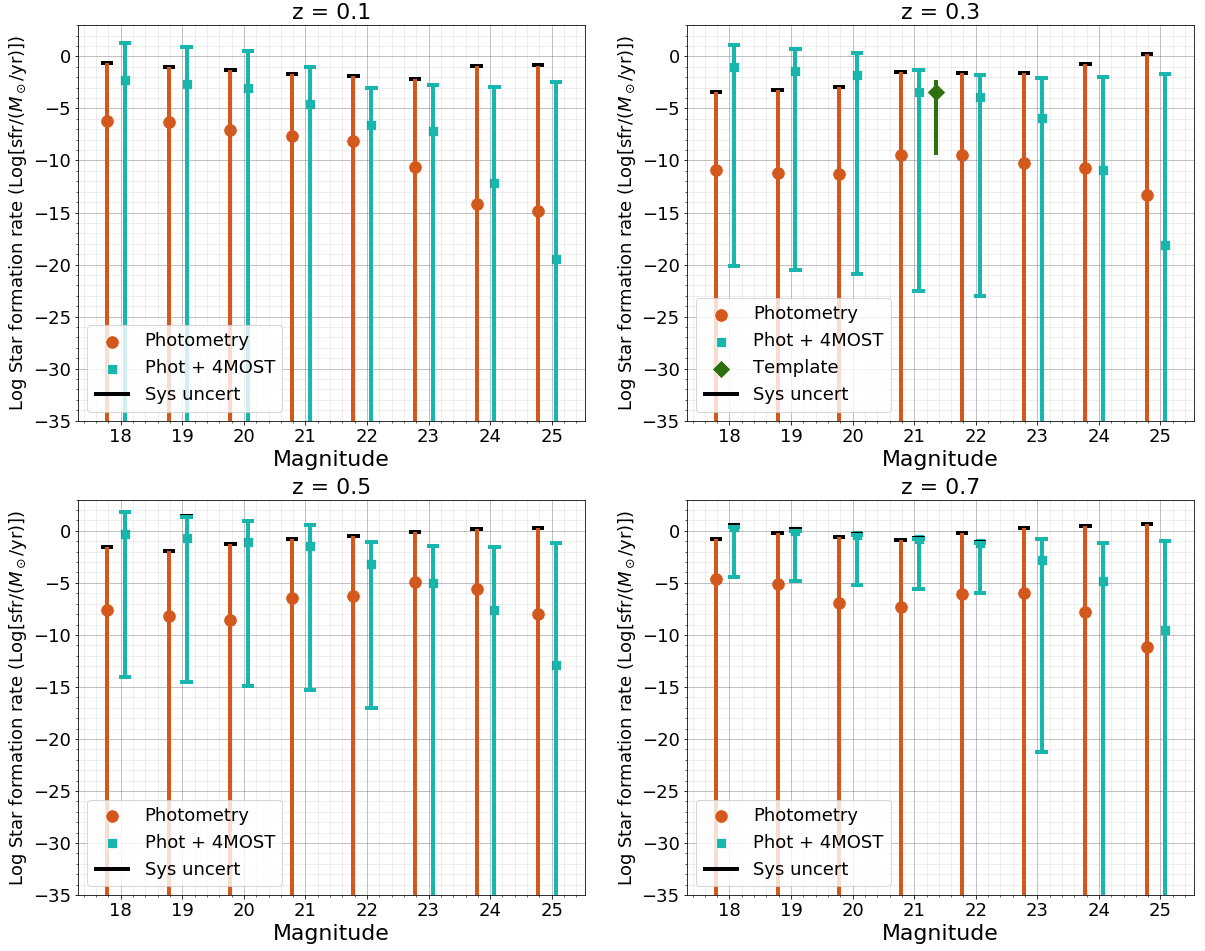}
    \caption{The simulated log Star Formation Rate (SFR) of an elliptical galaxy as a function of magnitude and redshift. The symbols and colours are the same as used in Figure \ref{fig: 4panel mass}. There is a clear improvement when 4MOST is used, compared with when photometry is used alone. This is true for all magnitudes brighter than 23, and redshifts 0.3, 0.5 and 0.7. However, at fainter magnitudes the uncertainty for 4MOST with photometry is a similar size to the photometry alone results. This is also true for redshift 0.1. A value for the improvement to star formation rate uncertainties is difficult to calculate as with photometry alone the lower value often hits FAST's lower limit of -99.}
    \label{fig: 4panel sfr plot}
\end{figure*}

\begin{figure*}
    \centering
    \includegraphics[scale=.38]{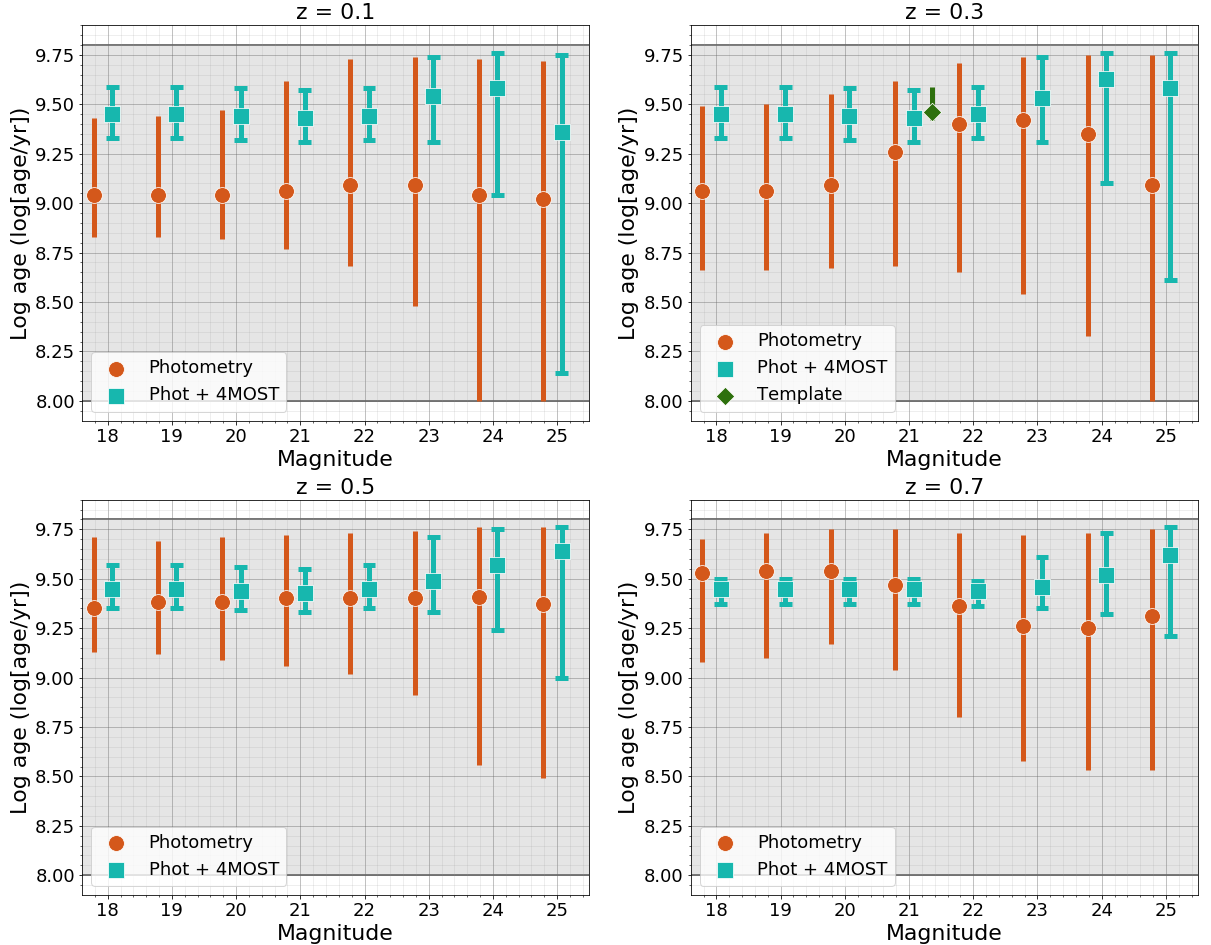}
    \caption{Simulated log age of of an elliptical galaxy as a function of magnitude and redshift. The precision increases drastically when 4MOST spectrum is used with photometry, compared with photometry alone. The limits that the age value can be found between by FAST are shown by the grey shaded region. The limits are set at 8 and 9.8 (log[age/yr]).}
    \label{fig: 4panel age}
\end{figure*}

\indent We are interested in the values FAST calculated for galaxy $M_\ast$, as this property has been found to correlate with SN properties. The results for galaxy $M_\ast$ can be seen in Figure \ref{fig: 4panel mass} and and Table \ref{Mass table el} in the appendix. The results for the template spectrum, simulated photometry and simulated spectroscopy at magnitude 21 and redshift 0.3, agree within the 68 percentile confidence ranges reported by FAST. This gives us confidence in the fitting process. The results show that for galaxies with $r$ = 22 mag and brighter, combining Rubin photometry and 4MOST spectroscopy reduces the uncertainty around the best-fit measurement of a galaxy's $M_\ast$ by 49 -- 95\% than when using photometry alone. There is also an improvement at magnitudes 23 to 25, however the improvement is not as significant.
\\\indent At fixed redshift the best-fit log ($M_\ast$) drops linearly as the apparent magnitude of the fitted galaxy increases, as shown in Figure \ref{fig: 4panel mass}. This is expected as FAST models the brightness of a galaxy as being proportional to the $M_\ast$ of the galaxy for a given stellar population. The best-fit galaxy parameters are calculated based on a multi-dimensional search with each of FAST's galaxy parameters. We find that at brighter magnitudes ($r$ $\le$ 22) some of the mass uncertanties reported by FAST are exactly zero, which is unrealistic. We believe that this is because the space between models in galaxy parameter space contained within FAST is not fine enough for FAST to find another model, within the 68\% confidence of the best-fit mass, with the small uncertainties provided by 4MOST.
\\\indent The results for the Sc galaxy can be seen in Figure \ref{fig: 4panel mass sc} and Table \ref{Mass table sc} (in the appendix). The results are comparable with the elliptical galaxy. Phot + 4MOST reduces the uncertainty for magnitudes $\le$ 22 by 79 -- 95\%, whilst at magnitudes fainter than 22 the uncertainty is reduced by 8 -- 68\%. The slope of the $M_\ast$ for each redshift is different for a Sc galaxy than the slope for the elliptical galaxy. This is due to other parameters within FAST finding different values for each galaxy, thus changing the mass slopes by a small amount.
\\\indent A recent publication by \citet{Pacifici22} reported on a systematic uncertainty when using SED fitting to measure galaxy properties. They reported on a systematic uncertainty which accounts for differences in SED fitting, as well as differences in model parameter assumptions. They explored a wide range of SED fitting codes, but did not analyse FAST. We used their median systematic values over all SED codes for mass (0.12 dex), star formation rate (0.27 dex) and V-band extinction (0.27 dex). These systematic values were added in quadrature to our uncertainties calculated from FAST. We have presented these results alongside our own. If a systematic uncertainty specifically for FAST is found in future studies then we can implement it. 
\\\indent Figures \ref{fig: circle el} and \ref{fig: circle sc} shows the uncertainty measurements on log ($M_\ast$) for each of the simulated magnitudes and redshift for the elliptical and Sc respectively. At all simulated redshifts and magnitudes there is an improvement to the precision of the measured host-galaxy $M_\ast$. We compare the uncertainties with and without the inclusion of spectroscopy. We see an improvement when spectroscopy is included at all simulated magnitudes and redshifts. A dashed black line across the plot represents the locus of points where $\log (\rm{mass}/M_{\odot}) = 10$. This was found using Figure \ref{fig: 4panel mass} and reading off the magnitude and redshift of a $10^{10}$ $M_{\odot}$ galaxy on each subplot. Note that the photometry values are calculated using the 10-year LSST depth. 4MOST spectroscopy will produce considerably better results than the intermediate photometry, until LSST reaches the 10th year of the survey. 
\\\indent In addition to the improvement to the precision of host-galaxy mass, all other galaxy parameters saw an improvement for phot + 4MOST compared with photometry alone. The results for an elliptical host-galaxy star formation rate can be seen in Figure \ref{fig: 4panel sfr plot}. There is a clear improvement for all magnitudes brighter than 23, and for redshifts 0.3, 0.5 and 0.7. However, at fainter magnitudes the uncertainty for phot + 4MOST is a similar size to the phot results. This is also true for redshift 0.1. A value for the improvement to star formation rate uncertainties is difficult to calculate because for photometry alone the lower value often hits FAST's lower limit of -99. At redshift 0.1 the uncertainty for phot + 4MOST is unconstrained due to a lack of models matching our galaxy in this parameter space. For the Sc galaxy we found that FAST had difficulties measuring the star formation rate from emission lines. 
\\\indent Uncertainties in age see improvements at all simulated redshifts and magnitudes. The results for this can be seen in Figure \ref{fig: 4panel age}. For phot + 4MOST, the uncertainty on the measurement of age was reduced by 56 -- 86\% for magnitudes $\le$ 22, while at magnitudes fainter than 22 the uncertainty was reduced by 6 -- 77\%. The improvements to age could prove to be a crucial detail that 4MOST spectra will be able to provide when calculating galaxy properties. Stellar absorption features are an indication of age, most notably the hydrogen Balmer lines \citep{Serra07}, which photometry is not able to capture \citep{Salim07}.
\\\indent Star formation rate and host-galaxy mass are the most commonly used parameters when making corrections in SNe cosmology. However all of the galaxy properties studied here may be of interest in the future because it is still unknown which galaxy parameters drive the correlations with SN properties. The work of \citet{Gupta11} found that after light curve corrections overluminous SNe Ia tend to occur in older stellar populations. The work of \citet{Rose21} found including age as a parameter improved the ability to standardise Type Ia SNe. The results of the study of the remaining galaxy properties calculated by FAST can be seen in the Appendix. The improved precision of galaxy property measurements could prove to be useful in the future at breaking degeneracies for SNe host-galaxy properties, such as age and metallicity \citep{Worthey94, Walcher11}. The key addition that 4MOST will provide is a large quantity of spectra with this level of quality. One limitation of our results is that FAST does not capture all of the information that a spectrum contains, such as emission lines. TiDES will be able to make direct measurements of emission lines, without fitting templates. For example, the H-alpha line will provide a further indication of star formation rate.

\section{Discussion} \label{Discussion}

\begin{figure} 
    \centering
    \includegraphics[scale=.32]{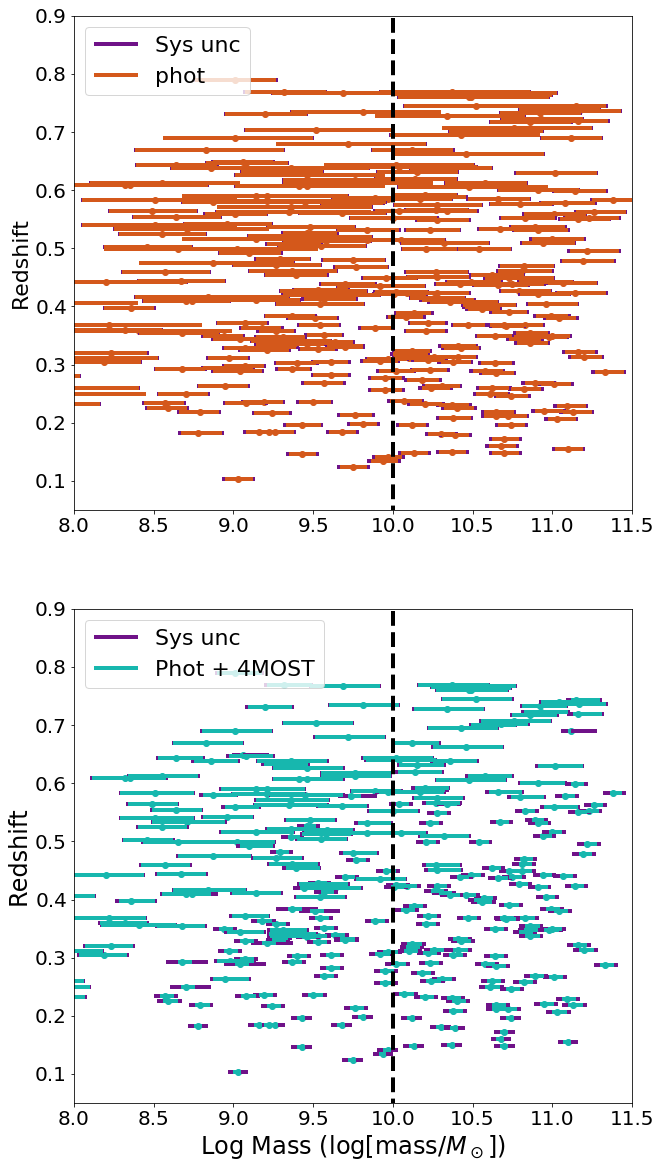}
    \caption{The mass-redshift distribution graph is created with our uncertainties. The SNe were taken from the Supernova Legacy Survey third year data set and the DES 3 year data. Each SN was assigned an uncertainty calculated from phot (in the top panel) and an uncertainty calculated from phot + 4MOST (in the bottom panel). Both plots have a black dashed line representing $10^{10}$ ($M_{\odot}$). There are 310 SNe plotted. The uncertainty values from the photometry lead to 35 SNe crossing the dashed line. Whilst, the uncertainty values from phot + 4MOST lead to 13 SNe crossing the dashed line.}
    \label{mass step graph}
\end{figure}

\indent Our work has shown that SNe host-galaxy $M_\ast$ can be measured more precisely when 4MOST and the Vera Rubin Observatory are used together, compared with the Rubin observatory alone. We expect to be able to improve the mass step correction by reducing the uncertainty as to which side of the $10^{10}M_\odot$ line a SN host galaxy belongs. 
\\\indent To quantify the extent of the improvement we have used the Supernova Legacy Survey third year data set \citep{Balland09}, used by \citet{Sullivan2010}, and the DES 3 year data \citep{Smith2020} to create a mass redshift distribution. While newer data sets could also have been used, these two data sets contain host-galaxy magnitudes and are sufficient to demonstrate our improvements in estimates of the host-galaxy mass. The Supernova Legacy Survey and DES 3 year data was readily accessible and large enough to demonstrate the effect of our results. \citeauthor{Balland09} report magnitudes of each of their SNe host-galaxies in $i$-band. Since our uncertainties correspond to $r$-band magnitudes, a conversion was required. To calculate the conversion, we used the integrated fluxes of our shifted target spectra through the $r$ and $i$ pass band (see Section \ref{Making phot}) to determine the colour as a function of redshift. We did this for the elliptical galaxy template, used in the analysis presented up to this point, and also a Sc galaxy template. For each SNe in \citeauthor{Balland09} we assigned a host-galaxy type at random, weighted by the rates at which SNe Ia appear in elliptical and Sc galaxies, as provided by \citet{Hakobyan12}. The $i$-band magnitude for the \citeauthor{Balland09} SNe host galaxy are converted to $r$-band using the colours calculated for the assigned galaxy type. The $r$-band magnitude of each galaxy was then rounded to the nearest integer value. The DES data provided $r$-band magnitudes so did not need any additional work. Each host galaxy was assigned mass uncertainties corresponding to its redshift and $r$-band magnitude bins, as previously shown in Figures \ref{fig: 4panel mass} and \ref{fig: 4panel mass sc}. Mass uncertainties were assigned for phot and phot + 4MOST, to enable a comparison. Host galaxies with redshift of 0.8 or greater were removed and any galaxies with $r$ band magnitudes greater than 25.5 were also removed. Host galaxies at higher redshifts and fainter magnitudes could not be used as the uncertainties start to increase rapidly for both LSST and 4MOST. A separate study of deep fields would be needed to determine the performance of our techniques for the faintest host galaxies. 
\\\indent After the cuts, we were left with a sample of 310 host galaxies. The results can be seen in Figure \ref{mass step graph}, where the top panel shows the host galaxies with the uncertainty produced from the photometry. The bottom panel shows the host-galaxies with the uncertainty produced by phot + 4MOST. The black dashed line on both plots is $10^{10}$ (mass/$M_{\odot}$), the divide for the mass step. With the photometry uncertainty values, 35 of the host-galaxies cross the mass step line, making it unclear which correction term would need to be applied. With the uncertainty values from phot + 4MOST, 13 host-galaxies cross the mass step line. This results in a 7\% improvement of SNe having the true correction applied. This shows that the uncertainty produced by phot + 4MOST would lead to a more accurate correction being applied than if only the photometry is used. It should be noted that we have assumed all host galaxies have spectra observed by 4MOST. The TiDES-Hosts sub-survey aims to capture spectra of SNe host galaxies, for which live observations were not captured \citet{Tides2019}. This will result in a significant number of SN host galaxies having spectra, which we will be able to select for when using results from 4MOST.
\\\indent We find on average our photometry mass uncertainty ranges are larger than those observed by \citet{Sullivan2010}. This is to be expected as \citeauthor{Sullivan2010} incorporates near-infrared filters (J, H, $K_s$) with photometry to measure a galaxy's mass. At magnitude 20 the SNLS data has an average mass uncertainty size of 0.12 dex, whilst our photometry uncertainty has a size of 0.19 dex. The phot + 4MOST uncertainty for magnitude 20 has a size of 0.08 dex. This further shows the improvement that 4MOST will enable. A similar trend occurs for magnitudes 21 and 22, with phot + 4MOST having a smaller uncertainty range than SNLS and phot only having a larger range. However, at magnitude 23 and beyond the 4MOST uncertainties become larger than the SNLS values (SNLS: 0.1 dex, phot: 0.68 dex, phot + 4MOST: 0.34 dex). Once again this is expected, as magnitude 23 is when the signal-to-noise for 4MOST becomes larger.
\\\indent A recent study by \cite{Galbany22} discussed aperture corrections to counteract effects caused by fixed-aperture fibre spectroscopy on host-galaxy correlations, such as the mass step. In this work we have effectively assumed that the magnitude of our host galaxies corresponds to the light which entered a 4MOST fibre. Using the galaxy catalogue published by \citet{Karachentsev04} we randomly selected several galaxies to calculate their size. We calculated a range of sizes between 2.48 arcseconds and 53.10 arcseconds. We previously stated that each 4MOST fibre will have a diameter of 1.45 arcseconds. As we selected objects randomly, there may be smaller and larger galaxies within the catalogue. This catalogue of galaxies only contained objects at redshift < 0.02. When we have access to real images from LSST we will be able to calculate the amount of light going down each fibre, and therefore apply aperture corrections.

\subsection{Analysing the effect of our work to cosmology} \label{Cosmology effect}
\begin{figure} 
    \centering
    \includegraphics[scale=.37]{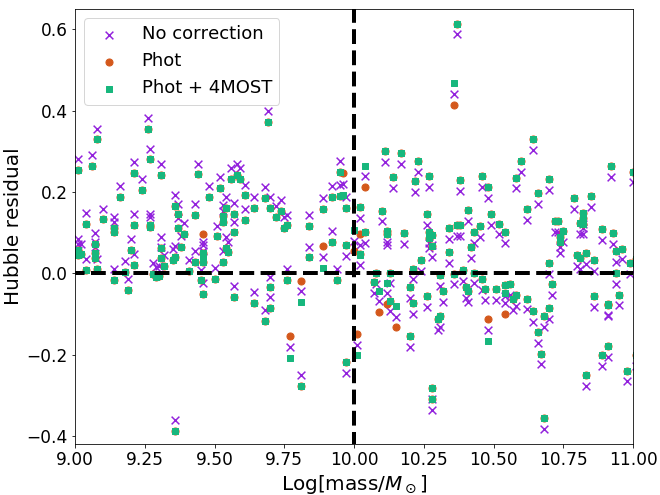}
    \caption{A diagram to illustrate the effect of the improved precision of galaxy stellar mass on Hubble residual. For each object in our selected sample three corrections are plotted, based on the stellar mass of the SN host galaxy. The purple crosses indicate no correction, the orange circles represent corrections assigned from photometric mass estimates, and the blue squares show corrections assigned from photometric + 4MOST mass estimates. In most cases, the corrections based on phot and phot + 4MOST are the same. The orange circles are visible when different corrections have been applied, otherwise the blue squares obscure them. As expected, near the $10^{10} M_\odot$ line it is more common to have disagreements between the correction terms applied. See  Section \ref{Cosmology effect} for more detail.}
    \label{Hubble correction graph}
\end{figure}

We now estimate the impact of our results on measurements of cosmological parameters. We investigated this by exploring the impact our results would have on the measurement of $w$. Measurements of $w$ are sensitive to small changes in distance modulus; a 2\% change in w corresponds to a change in distance modulus of only 2.6 millimagnitudes at $z$ = 0.3 (assuming a flat $\Lambda$CDM cosmology and all other parameters remain constant). We aim to determine how much our changes of mass uncertainties can improve scatter around the best-fit of the distance-redshift relation. We parameterise this using the root mean squared (RMS) scatter of the data points around the Hubble residual = 0 line.
\\\indent We calculated the Hubble residual of the corresponding SN for each of our host galaxies in Figure \ref{mass step graph} which appear in the Pantheon compilation \citep{2022Scolnic}, leaving us with 286 objects. The Hubble residual was calculated using a modified version of the Tripp formula \citep{Tripp98}, presented by \citet{Brout2019}, as shown as follows:

\begin{eqnarray}
\mathrm{Hubble\:Residual} = m_{\rm B, fix} - \mathrm{distance\:modulus} (z)\\
\mathrm{Where}\:\:m_{\rm B, fix} = (m_B\:+\:\alpha x_1\:-\:\beta c)\:-\:M_{B, SNIa}
\end{eqnarray}

where $m_B$ is the peak apparent  magnitude in the rest frame B-band. $M_{B, SNIa}$ is the mean absolute magnitude. This value was not provided in the Pantheon+ paper so we calculated it from the weighted average of the m\_b\_corr column (from Pantheon) minus the distance modulus, weighted with the MU\_SHOES\_ERR\_DIAG column (from Pantheon). $\alpha$ and $\beta$ are correlation coefficients of $x_1$ and $c$. The values for $m_B$, $x_1$ and $c$ were taken from the Pantheon dataset. The values for $\alpha$ and $\beta$ were set as 0.1533 and 3.44, respectively (taken from \citet{2022Brout}). 
The distance modulus was calculated using the distmod function in the \textsc{Python} module \textsc{Astropy}. 
We then calculated the probability of an object appearing either side of the $10^{10}$ (mass/$M_{\odot}$) line using Gaussian uncertainties, assuming our 1$\sigma$ values from FAST, centered on our best-fit $M_\ast$ values. This was repeated using our photometry uncertainties, and again for the phot + 4MOST uncertainties. 
We used the distance of each galaxy from the $10^{10}$ (mass/$M_{\odot}$) line and a Gaussian probability distribution to assign a probability of each host galaxy crossing the line. Each galaxy was then randomly assigned a right/wrong classification term, weighted by the probabilities, for a galaxy to be given the right or wrong correction term. The host galaxies with $M_\ast$ $< 10^{10} M_\odot$ were assigned a correction of $-0.0265$ mag (taken from \citeauthor{2022Brout}) whilst galaxies with $M_\ast$ $>= 10^{10} M_\odot$ were assigned a correction of $0.0265$ mag. This occurred for galaxies which were assigned the "right" correction term, else they were given the opposite correction value. Note that for simplicity we assumed a step function, whereas  \citeauthor{2022Brout} assumed an exponential, as described in \citet{Brout21}, which is almost identical.
\\\indent This is illustrated in Figure \ref{Hubble correction graph}. As expected only those host galaxies closest to the $10^{10} M_\odot$ line are effected. When the phot uncertainty is used to assign the mass correction, more points are assigned the wrong correction than when the phot + 4MOST uncertainties are used. In the figure, when the two corrections disagree the orange circles can be seen. When the corrections agree, the orange circles are obscured by the blue squares.

\indent To quantify the effect of the different corrections we calculated the RMS around the Hubble residual = 0 line. With no mass correction applied an RMS of 0.1752 mag was measured. When the mass correction calculated from the phot uncertainties were applied an RMS of 0.1720 mag was calculated. Finally, when the phot + 4MOST uncertainties were used to apply the mass correction, an RMS of 0.1715 mag was calculated. This is only a small improvement but represents the minimum improvement that 4MOST will be able to achieve to the mass correction. The SNe in our sample were taken from DES and SNLS, which observed each SN host galaxy with photometry. Hence, there could be ingrained uncertainty as to which side of the $10^{10} M_\odot$ mass line the host galaxy should be. To account for this we altered the way the mass correction is applied. As before, we assigned a right/wrong binary classification to decide if a galaxy should be given the right or wrong correction. We again used the Gaussian errors from phot and phot + 4MOST to assign the right/wrong classification. We then applied a correction based on the sign of the Hubble residual. All galaxies with a Hubble residual $>$ 0 were given a correction of $-0.0265$ mag and all galaxies with a Hubble residual $<$ 0 were assigned a correction of $0.0265$ mag. The same as the mass-based correction, galaxies assigned with the "right" correction term were given these corrections, else they were given the opposite correction value. The RMS was calculated again and found to be 0.1577 mag when the correction was applied using photometry uncertainties. The RMS was found to be 0.1566 mag when phot + 4MOST uncertainties were used. This represents the best improvement 4MOST will be able to accomplish. All of the RMS results are summarised in Table \ref{Table RMS}. We estimate the improvement to the RMS is between 0.0005 - 0.001 mag, which in the best case scenario would result in a 2\% improvement to the measurement uncertainty of $w$. We recognise that this is not significant to zeroth order, when calculating the mass step correction in this way. However, it is possible that such effects might be important when considering changes in $w$ with redshift (for example in $w_0 - w_a$ models), due to evolution of the host-galaxy population with redshift. We have not studied this effect.
\\\indent The results from our research will also be used to understand whether improved galaxy properties can aid classification of SNe Ia. To do this, further research will be undertaken using the machine learning algorithm SuperNNova \citep{Muller19} and a simulated catalogue of SNe light curves, with host-galaxy $M_\ast$, developed by \citet{Vincenzi19}. In addition, the results from our research can be improved with the addition of data from future instruments. ESA's \textit{Euclid} satellite \citep{Laureijs11} will provide near-infrared photometry for billions of galaxies over a wide area of the sky, and is due to launch in 2023. The \textit{ULTRASAT} satellite will provide the first wide-field ultraviolet time-domain sky survey and is due to launch in 2026 \citep{Ben-Ami22}. We expect using Euclid NIR photometry and ULTRASAT UV images, together with 4MOST and LSST data, will significantly improve the estimation of stellar mass and other galaxy properties. 

\setlength{\extrarowheight}{3pt}
\begin{table}
\centering
\begin{tabular}{|c|c|c|}
\hline
Correction & RMS Hubble & RMS stellar\\
        &  residual (mag) &  mass (mag)\\
\hline
None            & 0.1752  & 0.1752\\
Phot            & 0.1720  & 0.1577  \\
Phot + 4MOST    & 0.1715  & 0.1566  \\
\hline
\end{tabular}
\caption{The calculated RMS scatter values from 3 corrections: no correction, using the phot uncertainty and using phot + 4MOST uncertainty. The first column was calculated by measuring the scatter around the Hubble residual = 0 line, when a correction is applied based on which side of the $10^{10} M_\odot$ mass line a SN's host galaxy appears. These values are the minimum improvement which can be achieved. The second column is similar, but has a correction applied based on the sign of the Hubble residual, which is an attempt to account for ingrained uncertainty in the DES and SNLS host measurements. The scatter around the Hubble residual = 0 line is again measured. The second column also represents the maximum possible improvement. From these values, we estimate the improvement to the RMS is between
0.0005 - 0.001 mag.}
\label{Table RMS}
\end{table}

\section{Conclusions} \label{Summary}
\indent We have shown that 4MOST can work in conjunction with imaging telescopes, such as the Vera Rubin observatory, to calculate galaxy properties more precisely than those derived from photometry alone. This was an expected result which we have now quantified. Our results are summarised as follows:
\begin{itemize}
  \item For elliptical galaxies with brighter magnitudes ($r$ $\le$ 22) the uncertainty for a galaxy's log ($M_\ast$) is 49 -- 95\% smaller when spectroscopy is used with photometry, compared with that derived from photometry alone. The range of improvements depends on the magnitude and redshift.
  \item At fainter magnitudes ($r$ $\ge$ 23) the $M_\ast$ uncertainty is 24 -- 71\% smaller for phot + 4MOST, compared to photometry alone. The range depends on magnitude and redshift, with more improvements seen for brighter galaxies.
  \item We see similar improvements to the precision of Sc host-galaxy masses, when adding 4MOST spectroscopy.
  \item We applied our uncertainties derived from FAST to real SNe host-galaxy masses. The smaller uncertainties produced when adding 4MOST spectroscopy make it easier to distinguish which side of the $10^{10}$ ($M_{\odot}$) line a host galaxy falls. This has implications for applying the correct mass step corrections in cosmological analysis.
  \item Other galaxy properties see a significant improvement in uncertainties, including: star formation rate, age, V-band extinction, metallicity, specific star formation rate and star formation timescale.
  \item Whilst there is not as significant an improvement to the precision of a galaxy's metallicity, there is still an improvement at most simulated magnitudes and redshifts.
\end{itemize}

\indent The correlations between SNe Ia peak brightness and host-galaxy properties are one of the main systematic effects in SNe Ia cosmology. The improved galaxy property measurements from 4MOST and Rubin have the potential to improve the corrections used in SN cosmology. We have chosen to focus on the host-galaxy mass and the implications to the mass-step, but a deeper investigation could be carried out with all of the galaxy properties. The galaxy-property precision improvements come from the fact that spectra contain significantly more information than photometry, which leads to the breaking of degeneracies. We expect that information in the host-galaxy spectrum will also aid photometric classification of transients. With 4MOST beginning operations in late 2024 and the Vera Rubin Observatory beginning operations in 2025, we are about to enter an exciting period of SNe study.

\section*{Acknowledgements}
JD acknowledges the Science and Technology Facilities Council Data Science studentship and funding of training through the STFC 4IR Centre for Doctoral Training. IH acknowledges support for this work from STFC (consolidated grant numbers ST/R000514/1 and ST/V000713/1).  MN is supported by the European Research Council (ERC) under the European Union’s Horizon 2020 research and innovation programme (grant agreement No.~948381) and by funding from the UK Space Agency. BFR acknowledges support for this work by MNiSW grant DIR/WK/2018/12. This work has made use of the \textsc{python} adaptation of the FAST software developed by Corentin Schreiber, as well as documentation on FAST by James Aird. This work has made use of the development effort of 4MOST, an instrument being constructed by the 4MOST Consortium for the European Southern Observatory.

\section*{Data Availability}
The data used to create Figures \ref{fig: 4panel mass}, \ref{fig: 4panel mass sc} \ref{fig: 4panel sfr plot}, \ref{fig: 4panel age}, \ref{fig: 4panel av},  \ref{fig: 4panel tau}, \ref{fig: 4panel SSFR} and \ref{fig: 4panel metal} is available, as well as the results obtained for the Sc galaxy, at Lancaster University's data archive \url{https://doi.org/10.17635/lancaster/researchdata/621}.


\bibliographystyle{mnras}
\bibliography{references} 

\begin{thebibliography}{}
\makeatletter
\relax
\def\mn@urlcharsother{\let\do\@makeother \do\$\do\&\do\#\do\^\do\_\do\%\do\~}
\def\mn@doi{\begingroup\mn@urlcharsother \@ifnextchar [ {\mn@doi@}
  {\mn@doi@[]}}
\def\mn@doi@[#1]#2{\def\@tempa{#1}\ifx\@tempa\@empty \href
  {http://dx.doi.org/#2} {doi:#2}\else \href {http://dx.doi.org/#2} {#1}\fi
  \endgroup}
\def\mn@eprint#1#2{\mn@eprint@#1:#2::\@nil}
\def\mn@eprint@arXiv#1{\href {http://arxiv.org/abs/#1} {{\tt arXiv:#1}}}
\def\mn@eprint@dblp#1{\href {http://dblp.uni-trier.de/rec/bibtex/#1.xml}
  {dblp:#1}}
\def\mn@eprint@#1:#2:#3:#4\@nil{\def\@tempa {#1}\def\@tempb {#2}\def\@tempc
  {#3}\ifx \@tempc \@empty \let \@tempc \@tempb \let \@tempb \@tempa \fi \ifx
  \@tempb \@empty \def\@tempb {arXiv}\fi \@ifundefined
  {mn@eprint@\@tempb}{\@tempb:\@tempc}{\expandafter \expandafter \csname
  mn@eprint@\@tempb\endcsname \expandafter{\@tempc}}}

\bibitem[\protect\citeauthoryear{Alam et~al.,}{Alam et~al.}{2015}]{Alam15}
Alam S.,  et~al., 2015, \mn@doi [The Astrophysical Journal Supplement Series]
  {10.1088/0067-0049/219/1/12}, 219, 12

\bibitem[\protect\citeauthoryear{{Aldering} et~al.,}{{Aldering}
  et~al.}{2002}]{2002Aldering}
{Aldering} G.,  et~al., 2002, in {Tyson} J.~A.,  {Wolff} S.,  eds,  Society of
  Photo-Optical Instrumentation Engineers (SPIE) Conference Series Vol. 4836,
  Survey and Other Telescope Technologies and Discoveries. pp 61--72,
  \mn@doi{10.1117/12.458107}

\bibitem[\protect\citeauthoryear{{Balland} et~al.,}{{Balland}
  et~al.}{2009}]{Balland09}
{Balland} C.,  et~al., 2009, \mn@doi [\aap] {10.1051/0004-6361/200912246},
  \href {https://ui.adsabs.harvard.edu/abs/2009A&A...507...85B} {507, 85}

\bibitem[\protect\citeauthoryear{Ben-Ami et~al.,}{Ben-Ami
  et~al.}{2022}]{Ben-Ami22}
Ben-Ami S.,  et~al., 2022, The scientific payload of the Ultraviolet Transient
  Astronomy Satellite (ULTRASAT), \mn@doi{10.48550/ARXIV.2208.00159}, \url
  {https://arxiv.org/abs/2208.00159}

\bibitem[\protect\citeauthoryear{{Botticella} et~al.,}{{Botticella}
  et~al.}{2017}]{Botticella17}
{Botticella} M.~T.,  et~al., 2017, \mn@doi [\aap]
  {10.1051/0004-6361/201629432}, \href
  {https://ui.adsabs.harvard.edu/abs/2017A&A...598A..50B} {598, A50}

\bibitem[\protect\citeauthoryear{{Branch} \& {Tammann}}{{Branch} \&
  {Tammann}}{1992}]{92Branch}
{Branch} D.,  {Tammann} G.~A.,  1992, \mn@doi [\araa]
  {10.1146/annurev.aa.30.090192.002043}, \href
  {https://ui.adsabs.harvard.edu/abs/1992ARA&A..30..359B} {30, 359}

\bibitem[\protect\citeauthoryear{{Briday} et~al.,}{{Briday}
  et~al.}{2022}]{22Briday}
{Briday} M.,  et~al., 2022, \mn@doi [\aap] {10.1051/0004-6361/202141160}, \href
  {https://ui.adsabs.harvard.edu/abs/2022A&A...657A..22B} {657, A22}

\bibitem[\protect\citeauthoryear{Brout \& Scolnic}{Brout \&
  Scolnic}{2021}]{Brout21}
Brout D.,  Scolnic D.,  2021, \mn@doi [The Astrophysical Journal]
  {10.3847/1538-4357/abd69b}, 909, 26

\bibitem[\protect\citeauthoryear{{Brout} et~al.,}{{Brout}
  et~al.}{2019}]{Brout2019}
{Brout} D.,  et~al., 2019, \mn@doi [\apj] {10.3847/1538-4357/ab08a0}, \href
  {https://ui.adsabs.harvard.edu/abs/2019ApJ...874..150B} {874, 150}

\bibitem[\protect\citeauthoryear{{Brout} et~al.,}{{Brout}
  et~al.}{2022}]{2022Brout}
{Brout} D.,  et~al., 2022, \mn@doi [\apj] {10.3847/1538-4357/ac8e04}, \href
  {https://ui.adsabs.harvard.edu/abs/2022ApJ...938..110B} {938, 110}

\bibitem[\protect\citeauthoryear{{Bruzual} \& {Charlot}}{{Bruzual} \&
  {Charlot}}{2003}]{03Bruzual}
{Bruzual} G.,  {Charlot} S.,  2003, \mn@doi [\mnras]
  {10.1046/j.1365-8711.2003.06897.x}, \href
  {https://ui.adsabs.harvard.edu/abs/2003MNRAS.344.1000B} {344, 1000}

\bibitem[\protect\citeauthoryear{{Chabrier}}{{Chabrier}}{2003}]{03Chabrier}
{Chabrier} G.,  2003, \mn@doi [\pasp] {10.1086/376392}, \href
  {https://ui.adsabs.harvard.edu/abs/2003PASP..115..763C} {115, 763}

\bibitem[\protect\citeauthoryear{{Chen} et~al.,}{{Chen}
  et~al.}{2022}]{2022Chen}
{Chen} R.,  et~al., 2022, \mn@doi [\apj] {10.3847/1538-4357/ac8b82}, \href
  {https://ui.adsabs.harvard.edu/abs/2022ApJ...938...62C} {938, 62}

\bibitem[\protect\citeauthoryear{Childress et~al.,}{Childress
  et~al.}{2013}]{Childress13}
Childress M.,  et~al., 2013, \mn@doi [The Astrophysical Journal]
  {10.1088/0004-637x/770/2/107}, 770, 107

\bibitem[\protect\citeauthoryear{{Conroy}}{{Conroy}}{2013}]{13Conroy}
{Conroy} C.,  2013, \mn@doi [\araa] {10.1146/annurev-astro-082812-141017},
  \href {https://ui.adsabs.harvard.edu/abs/2013ARA&A..51..393C} {51, 393}

\bibitem[\protect\citeauthoryear{{Dhawan}, {Thorp}, {Mandel}, {Ward},
  {Narayan}, {Jha}  \& {Chant}}{{Dhawan} et~al.}{2022}]{2022Dhawan}
{Dhawan} S.,  {Thorp} S.,  {Mandel} K.~S.,  {Ward} S.~M.,  {Narayan} G.,  {Jha}
  S.~W.,   {Chant} T.,  2022, arXiv e-prints, \href
  {https://ui.adsabs.harvard.edu/abs/2022arXiv221107657D} {p. arXiv:2211.07657}

\bibitem[\protect\citeauthoryear{{Euclid Collaboration} et~al.,}{{Euclid
  Collaboration} et~al.}{2022}]{Humphrey22}
{Euclid Collaboration} et~al., 2022, \mn@doi [arXiv e-prints]
  {10.48550/arXiv.2209.13074}, \href
  {https://ui.adsabs.harvard.edu/abs/2022arXiv220913074E} {p. arXiv:2209.13074}

\bibitem[\protect\citeauthoryear{Foley \& Mandel}{Foley \&
  Mandel}{2013}]{Foley13}
Foley R.~J.,  Mandel K.,  2013, \mn@doi [The Astrophysical Journal]
  {10.1088/0004-637x/778/2/167}, 778, 167

\bibitem[\protect\citeauthoryear{{Gagliano}, {Narayan}, {Engel}, {Carrasco
  Kind}  \& {LSST Dark Energy Science Collaboration}}{{Gagliano}
  et~al.}{2021}]{Gagliano21}
{Gagliano} A.,  {Narayan} G.,  {Engel} A.,  {Carrasco Kind} M.,   {LSST Dark
  Energy Science Collaboration} 2021, \mn@doi [\apj]
  {10.3847/1538-4357/abd02b}, \href
  {https://ui.adsabs.harvard.edu/abs/2021ApJ...908..170G} {908, 170}

\bibitem[\protect\citeauthoryear{{Galbany} et~al.,}{{Galbany}
  et~al.}{2014}]{Galbany14}
{Galbany} L.,  et~al., 2014, \mn@doi [\aap] {10.1051/0004-6361/201424717},
  \href {https://ui.adsabs.harvard.edu/abs/2014A&A...572A..38G} {572, A38}

\bibitem[\protect\citeauthoryear{{Galbany} et~al.,}{{Galbany}
  et~al.}{2022}]{Galbany22}
{Galbany} L.,  et~al., 2022, \mn@doi [\aap] {10.1051/0004-6361/202141568},
  \href {https://ui.adsabs.harvard.edu/abs/2022A&A...659A..89G} {659, A89}

\bibitem[\protect\citeauthoryear{Gallagher, Garnavich, Caldwell, Kirshner, Jha,
  Li, Ganeshalingam  \& Filippenko}{Gallagher et~al.}{2008}]{Gallagher08}
Gallagher J.~S.,  Garnavich P.~M.,  Caldwell N.,  Kirshner R.~P.,  Jha S.~W.,
  Li W.,  Ganeshalingam M.,   Filippenko A.~V.,  2008, \mn@doi [The
  Astrophysical Journal] {10.1086/590659}, 685, 752

\bibitem[\protect\citeauthoryear{{Garnavich} et~al.,}{{Garnavich}
  et~al.}{1998}]{Garnavich98}
{Garnavich} P.~M.,  et~al., 1998, \mn@doi [\apj] {10.1086/306495}, \href
  {https://ui.adsabs.harvard.edu/abs/1998ApJ...509...74G} {509, 74}

\bibitem[\protect\citeauthoryear{{Graur}, {Bianco}, {Huang}, {Modjaz},
  {Shivvers}, {Filippenko}, {Li}  \& {Eldridge}}{{Graur}
  et~al.}{2017a}]{2017Graura}
{Graur} O.,  {Bianco} F.~B.,  {Huang} S.,  {Modjaz} M.,  {Shivvers} I.,
  {Filippenko} A.~V.,  {Li} W.,   {Eldridge} J.~J.,  2017a, \mn@doi [\apj]
  {10.3847/1538-4357/aa5eb8}, \href
  {https://ui.adsabs.harvard.edu/abs/2017ApJ...837..120G} {837, 120}

\bibitem[\protect\citeauthoryear{{Graur}, {Bianco}, {Modjaz}, {Shivvers},
  {Filippenko}, {Li}  \& {Smith}}{{Graur} et~al.}{2017b}]{2017Graurb}
{Graur} O.,  {Bianco} F.~B.,  {Modjaz} M.,  {Shivvers} I.,  {Filippenko} A.~V.,
   {Li} W.,   {Smith} N.,  2017b, \mn@doi [\apj] {10.3847/1538-4357/aa5eb7},
  \href {https://ui.adsabs.harvard.edu/abs/2017ApJ...837..121G} {837, 121}

\bibitem[\protect\citeauthoryear{{Guiglion} et~al.,}{{Guiglion}
  et~al.}{2019a}]{Guiglion19}
{Guiglion} G.,  et~al., 2019a, \mn@doi [The Messenger]
  {10.18727/0722-6691/5119}, \href
  {https://ui.adsabs.harvard.edu/abs/2019Msngr.175...17G} {175, 17}

\bibitem[\protect\citeauthoryear{Guiglion et~al.,}{Guiglion
  et~al.}{2019b}]{4MOST_obs}
Guiglion G.,  et~al., 2019b, \mn@doi [Published in The Messenger vol. 175]
  {10.18727/0722-6691/5119}, pp. 17-21, March 2019.

\bibitem[\protect\citeauthoryear{{Gupta} et~al.,}{{Gupta}
  et~al.}{2011}]{Gupta11}
{Gupta} R.~R.,  et~al., 2011, \mn@doi [\apj] {10.1088/0004-637X/740/2/92},
  \href {https://ui.adsabs.harvard.edu/abs/2011ApJ...740...92G} {740, 92}

\bibitem[\protect\citeauthoryear{{Hakobyan}, {Adibekyan, V. Zh.}, {Aramyan, L.
  S.}, {Petrosian, A. R.}, {Gomes, J. M.}, {Mamon, G. A.}, {Kunth, D.}  \&
  {Turatto, M.}}{{Hakobyan} et~al.}{2012}]{Hakobyan12}
{Hakobyan} A.~A.,  {Adibekyan, V. Zh.} {Aramyan, L. S.} {Petrosian, A. R.}
  {Gomes, J. M.} {Mamon, G. A.} {Kunth, D.}  {Turatto, M.} 2012, \mn@doi [A\&A]
  {10.1051/0004-6361/201219541}, 544, A81

\bibitem[\protect\citeauthoryear{{Hamuy}, {Trager}, {Pinto}, {Phillips},
  {Schommer}, {Ivanov}  \& {Suntzeff}}{{Hamuy} et~al.}{2001}]{Hamuy01}
{Hamuy} M.,  {Trager} S.~C.,  {Pinto} P.~A.,  {Phillips} M.~M.,  {Schommer}
  R.~A.,  {Ivanov} V.,   {Suntzeff} N.~B.,  2001, \mn@doi [\aj]
  {10.1086/324251}, \href
  {https://ui.adsabs.harvard.edu/abs/2001AJ....122.3506H} {122, 3506}

\bibitem[\protect\citeauthoryear{{Holwerda}}{{Holwerda}}{2008}]{Holwerda08}
{Holwerda} B.~W.,  2008, \mn@doi [\mnras] {10.1111/j.1365-2966.2008.13050.x},
  \href {https://ui.adsabs.harvard.edu/abs/2008MNRAS.386..475H} {386, 475}

\bibitem[\protect\citeauthoryear{{Holwerda}, {Reynolds}, {Smith}  \&
  {Kraan-Korteweg}}{{Holwerda} et~al.}{2015}]{Holwerda15}
{Holwerda} B.~W.,  {Reynolds} A.,  {Smith} M.,   {Kraan-Korteweg} R.~C.,  2015,
  \mn@doi [\mnras] {10.1093/mnras/stu2345}, \href
  {https://ui.adsabs.harvard.edu/abs/2015MNRAS.446.3768H} {446, 3768}

\bibitem[\protect\citeauthoryear{Ivezi{\'c} \& the LSST
  Science~Collaboration}{Ivezi{\'c} \& the LSST
  Science~Collaboration}{2013}]{LSSTDoc}
Ivezi{\'c} {\v{Z}.}.,  the LSST Science~Collaboration 2013, LSST Science
  Requirements Document, \url {http://ls.st/LPM-17}

\bibitem[\protect\citeauthoryear{{Ivezi{\'c}} et~al.,}{{Ivezi{\'c}}
  et~al.}{2019}]{Ivezic2019}
{Ivezi{\'c}} {\v{Z}}.,  et~al., 2019, \mn@doi [\apj]
  {10.3847/1538-4357/ab042c}, \href
  {https://ui.adsabs.harvard.edu/abs/2019ApJ...873..111I} {873, 111}

\bibitem[\protect\citeauthoryear{{Jha}, {Riess}  \& {Kirshner}}{{Jha}
  et~al.}{2007}]{Jha07}
{Jha} S.,  {Riess} A.~G.,   {Kirshner} R.~P.,  2007, \mn@doi [\apj]
  {10.1086/512054}, \href
  {https://ui.adsabs.harvard.edu/abs/2007ApJ...659..122J} {659, 122}

\bibitem[\protect\citeauthoryear{{Jones}, {Stanway}  \& {Carnall}}{{Jones}
  et~al.}{2022}]{2022Jones}
{Jones} G.~T.,  {Stanway} E.~R.,   {Carnall} A.~C.,  2022, \mn@doi [\mnras]
  {10.1093/mnras/stac1667}, \href
  {https://ui.adsabs.harvard.edu/abs/2022MNRAS.514.5706J} {514, 5706}

\bibitem[\protect\citeauthoryear{{Karachentsev}, {Karachentseva}, {Huchtmeier}
  \& {Makarov}}{{Karachentsev} et~al.}{2004}]{Karachentsev04}
{Karachentsev} I.~D.,  {Karachentseva} V.~E.,  {Huchtmeier} W.~K.,   {Makarov}
  D.~I.,  2004, \mn@doi [\aj] {10.1086/382905}, \href
  {https://ui.adsabs.harvard.edu/abs/2004AJ....127.2031K} {127, 2031}

\bibitem[\protect\citeauthoryear{{Kelly}, {Hicken}, {Burke}, {Mandel}  \&
  {Kirshner}}{{Kelly} et~al.}{2010}]{Kelly10}
{Kelly} P.~L.,  {Hicken} M.,  {Burke} D.~L.,  {Mandel} K.~S.,   {Kirshner}
  R.~P.,  2010, \mn@doi [\apj] {10.1088/0004-637X/715/2/743}, \href
  {https://ui.adsabs.harvard.edu/abs/2010ApJ...715..743K} {715, 743}

\bibitem[\protect\citeauthoryear{Kelsey et~al.,}{Kelsey
  et~al.}{2020}]{Kelsey20}
Kelsey L.,  et~al., 2020, \mn@doi [Monthly Notices of the Royal Astronomical
  Society] {10.1093/mnras/staa3924}, 501, 4861

\bibitem[\protect\citeauthoryear{{Kinney}, {Calzetti}, {Bohlin}, {McQuade},
  {Storchi-Bergmann}  \& {Schmitt}}{{Kinney} et~al.}{1996}]{96Kinney}
{Kinney} A.~L.,  {Calzetti} D.,  {Bohlin} R.~C.,  {McQuade} K.,
  {Storchi-Bergmann} T.,   {Schmitt} H.~R.,  1996, \mn@doi [\apj]
  {10.1086/177583}, \href
  {https://ui.adsabs.harvard.edu/abs/1996ApJ...467...38K} {467, 38}

\bibitem[\protect\citeauthoryear{{Kokusho}, {Kaneda}, {Bureau}, {Suzuki},
  {Murata}, {Kondo}  \& {Yamagishi}}{{Kokusho} et~al.}{2017}]{2017Kokusho}
{Kokusho} T.,  {Kaneda} H.,  {Bureau} M.,  {Suzuki} T.,  {Murata} K.,  {Kondo}
  A.,   {Yamagishi} M.,  2017, \mn@doi [\aap] {10.1051/0004-6361/201630158},
  \href {https://ui.adsabs.harvard.edu/abs/2017A&A...605A..74K} {605, A74}

\bibitem[\protect\citeauthoryear{{Kowalski} et~al.,}{{Kowalski}
  et~al.}{2008}]{08Kowalski}
{Kowalski} M.,  et~al., 2008, \mn@doi [\apj] {10.1086/589937}, \href
  {https://ui.adsabs.harvard.edu/abs/2008ApJ...686..749K} {686, 749}

\bibitem[\protect\citeauthoryear{{Kriek} \& {Conroy}}{{Kriek} \&
  {Conroy}}{2013}]{13Kriek}
{Kriek} M.,  {Conroy} C.,  2013, \mn@doi [\apjl] {10.1088/2041-8205/775/1/L16},
  \href {https://ui.adsabs.harvard.edu/abs/2013ApJ...775L..16K} {775, L16}

\bibitem[\protect\citeauthoryear{{Kriek}, {van Dokkum}, {Labb{\'e}}, {Franx},
  {Illingworth}, {Marchesini}  \& {Quadri}}{{Kriek} et~al.}{2009}]{Kriek2009}
{Kriek} M.,  {van Dokkum} P.~G.,  {Labb{\'e}} I.,  {Franx} M.,  {Illingworth}
  G.~D.,  {Marchesini} D.,   {Quadri} R.~F.,  2009, \mn@doi [\apj]
  {10.1088/0004-637X/700/1/221}, \href
  {https://ui.adsabs.harvard.edu/abs/2009ApJ...700..221K} {700, 221}

\bibitem[\protect\citeauthoryear{{LSST Science Collaboration} et~al.,}{{LSST
  Science Collaboration} et~al.}{2009}]{09LSST}
{LSST Science Collaboration} et~al., 2009, \mn@doi [arXiv e-prints]
  {10.48550/arXiv.0912.0201}, \href
  {https://ui.adsabs.harvard.edu/abs/2009arXiv0912.0201L} {p. arXiv:0912.0201}

\bibitem[\protect\citeauthoryear{{Lampeitl} et~al.,}{{Lampeitl}
  et~al.}{2010}]{Lampeitl2010}
{Lampeitl} H.,  et~al., 2010, \mn@doi [\apj] {10.1088/0004-637X/722/1/566},
  \href {https://ui.adsabs.harvard.edu/abs/2010ApJ...722..566L} {722, 566}

\bibitem[\protect\citeauthoryear{Laureijs et~al.,}{Laureijs
  et~al.}{2011}]{Laureijs11}
Laureijs R.,  et~al., 2011, Euclid Definition Study Report,
  \mn@doi{10.48550/ARXIV.1110.3193}, \url {https://arxiv.org/abs/1110.3193}

\bibitem[\protect\citeauthoryear{{Li} et~al.,}{{Li} et~al.}{2022}]{2022Li}
{Li} M.,  et~al., 2022, arXiv e-prints, \href
  {https://ui.adsabs.harvard.edu/abs/2022arXiv221101382L} {p. arXiv:2211.01382}

\bibitem[\protect\citeauthoryear{{Lidman} et~al.,}{{Lidman}
  et~al.}{2020}]{Lidman20}
{Lidman} C.,  et~al., 2020, \mn@doi [\mnras] {10.1093/mnras/staa1341}, \href
  {https://ui.adsabs.harvard.edu/abs/2020MNRAS.496...19L} {496, 19}

\bibitem[\protect\citeauthoryear{{Lower}, {Narayanan}, {Leja}, {Johnson},
  {Conroy}  \& {Dav{\'e}}}{{Lower} et~al.}{2020}]{2020Lower}
{Lower} S.,  {Narayanan} D.,  {Leja} J.,  {Johnson} B.~D.,  {Conroy} C.,
  {Dav{\'e}} R.,  2020, \mn@doi [\apj] {10.3847/1538-4357/abbfa7}, \href
  {https://ui.adsabs.harvard.edu/abs/2020ApJ...904...33L} {904, 33}

\bibitem[\protect\citeauthoryear{{Mannucci}, {Della Valle}, {Panagia},
  {Cappellaro}, {Cresci}, {Maiolino}, {Petrosian}  \& {Turatto}}{{Mannucci}
  et~al.}{2005}]{Mannucci05}
{Mannucci} F.,  {Della Valle} M.,  {Panagia} N.,  {Cappellaro} E.,  {Cresci}
  G.,  {Maiolino} R.,  {Petrosian} A.,   {Turatto} M.,  2005, \mn@doi [A\&A]
  {10.1051/0004-6361:20041411}, 433, 807

\bibitem[\protect\citeauthoryear{Marshall et~al.,}{Marshall
  et~al.}{2017}]{LSST_obs}
Marshall P.,  et~al., 2017, Lsst Science Collaborations Observing Strategy
  White Paper: "Science-Driven Optimization Of The Lsst Observing Strategy",
  \mn@doi{10.5281/ZENODO.842713}, \url {https://zenodo.org/record/842713}

\bibitem[\protect\citeauthoryear{{Modjaz} et~al.,}{{Modjaz}
  et~al.}{2020}]{Modjaz20}
{Modjaz} M.,  et~al., 2020, \mn@doi [\apj] {10.3847/1538-4357/ab4185}, \href
  {https://ui.adsabs.harvard.edu/abs/2020ApJ...892..153M} {892, 153}

\bibitem[\protect\citeauthoryear{{Morrissey} et~al.,}{{Morrissey}
  et~al.}{2007}]{2007Morrissey}
{Morrissey} P.,  et~al., 2007, \mn@doi [\apjs] {10.1086/520512}, \href
  {https://ui.adsabs.harvard.edu/abs/2007ApJS..173..682M} {173, 682}

\bibitem[\protect\citeauthoryear{Möller \& de Boissi{\`{e} }re}{Möller \&
  de~Boissi{\`{e} }re}{2019}]{Muller19}
Möller A.,  de Boissi{\`{e} }re T.,  2019, \mn@doi [Monthly Notices of the
  Royal Astronomical Society] {10.1093/mnras/stz3312}, 491, 4277

\bibitem[\protect\citeauthoryear{{Oemler} \& {Tinsley}}{{Oemler} \&
  {Tinsley}}{1979}]{Oemler79}
{Oemler} A. J.,  {Tinsley} B.~M.,  1979, \mn@doi [\aj] {10.1086/112502}, \href
  {https://ui.adsabs.harvard.edu/abs/1979AJ.....84..985O} {84, 985}

\bibitem[\protect\citeauthoryear{{Pacifici} et~al.,}{{Pacifici}
  et~al.}{2022}]{Pacifici22}
{Pacifici} C.,  et~al., 2022, \mn@doi [arXiv e-prints]
  {10.48550/arXiv.2212.01915}, \href
  {https://ui.adsabs.harvard.edu/abs/2022arXiv221201915P} {p. arXiv:2212.01915}

\bibitem[\protect\citeauthoryear{{Pan} et~al.,}{{Pan} et~al.}{2014}]{Pan14}
{Pan} Y.~C.,  et~al., 2014, \mn@doi [\mnras] {10.1093/mnras/stt2287}, \href
  {https://ui.adsabs.harvard.edu/abs/2014MNRAS.438.1391P} {438, 1391}

\bibitem[\protect\citeauthoryear{{Perlmutter} et~al.,}{{Perlmutter}
  et~al.}{1999}]{Perlmutter1999}
{Perlmutter} S.,  et~al., 1999, \mn@doi [\apj] {10.1086/307221}, \href
  {https://ui.adsabs.harvard.edu/abs/1999ApJ...517..565P} {517, 565}

\bibitem[\protect\citeauthoryear{{Riess} et~al.,}{{Riess}
  et~al.}{1998}]{Reiss1998}
{Riess} A.~G.,  et~al., 1998, \mn@doi [\aj] {10.1086/300499}, \href
  {https://ui.adsabs.harvard.edu/abs/1998AJ....116.1009R} {116, 1009}

\bibitem[\protect\citeauthoryear{Riess et~al.,}{Riess et~al.}{2004}]{04Riess}
Riess A.~G.,  et~al., 2004, \mn@doi [The Astrophysical Journal]
  {10.1086/383612}, 607, 665

\bibitem[\protect\citeauthoryear{{Riess} et~al.,}{{Riess}
  et~al.}{2007}]{07Riess}
{Riess} A.~G.,  et~al., 2007, \mn@doi [\apj] {10.1086/510378}, \href
  {https://ui.adsabs.harvard.edu/abs/2007ApJ...659...98R} {659, 98}

\bibitem[\protect\citeauthoryear{{Rigault} et~al.,}{{Rigault}
  et~al.}{2020}]{2020Rigault}
{Rigault} M.,  et~al., 2020, \mn@doi [\aap] {10.1051/0004-6361/201730404},
  \href {https://ui.adsabs.harvard.edu/abs/2020A&A...644A.176R} {644, A176}

\bibitem[\protect\citeauthoryear{Rose, Rubin, Strolger  \& Garnavich}{Rose
  et~al.}{2021}]{Rose21}
Rose B.~M.,  Rubin D.,  Strolger L.,   Garnavich P.~M.,  2021, \mn@doi [The
  Astrophysical Journal] {10.3847/1538-4357/abd550}, 909, 28

\bibitem[\protect\citeauthoryear{{Salim} et~al.,}{{Salim}
  et~al.}{2007}]{Salim07}
{Salim} S.,  et~al., 2007, \mn@doi [\apjs] {10.1086/519218}, \href
  {https://ui.adsabs.harvard.edu/abs/2007ApJS..173..267S} {173, 267}

\bibitem[\protect\citeauthoryear{Scolnic et~al.,}{Scolnic
  et~al.}{2018}]{Scolnic18}
Scolnic D.~M.,  et~al., 2018, \mn@doi [The Astrophysical Journal]
  {10.3847/1538-4357/aab9bb}, 859, 101

\bibitem[\protect\citeauthoryear{{Scolnic} et~al.,}{{Scolnic}
  et~al.}{2022}]{2022Scolnic}
{Scolnic} D.,  et~al., 2022, \mn@doi [\apj] {10.3847/1538-4357/ac8b7a}, \href
  {https://ui.adsabs.harvard.edu/abs/2022ApJ...938..113S} {938, 113}

\bibitem[\protect\citeauthoryear{{Serra} \& {Trager}}{{Serra} \&
  {Trager}}{2007}]{Serra07}
{Serra} P.,  {Trager} S.~C.,  2007, \mn@doi [\mnras]
  {10.1111/j.1365-2966.2006.11188.x}, \href
  {https://ui.adsabs.harvard.edu/abs/2007MNRAS.374..769S} {374, 769}

\bibitem[\protect\citeauthoryear{{Smith} et~al.,}{{Smith}
  et~al.}{2020}]{Smith2020}
{Smith} M.,  et~al., 2020, \mn@doi [\aj] {10.3847/1538-3881/abc01b}, \href
  {https://ui.adsabs.harvard.edu/abs/2020AJ....160..267S} {160, 267}

\bibitem[\protect\citeauthoryear{{Speagle}, {Steinhardt}, {Capak}  \&
  {Silverman}}{{Speagle} et~al.}{2014}]{2014Speagle}
{Speagle} J.~S.,  {Steinhardt} C.~L.,  {Capak} P.~L.,   {Silverman} J.~D.,
  2014, \mn@doi [\apjs] {10.1088/0067-0049/214/2/15}, \href
  {https://ui.adsabs.harvard.edu/abs/2014ApJS..214...15S} {214, 15}

\bibitem[\protect\citeauthoryear{{Spinrad}}{{Spinrad}}{1972}]{Spinrad72}
{Spinrad} H.,  1972, \mn@doi [\apj] {10.1086/151299}, \href
  {https://ui.adsabs.harvard.edu/abs/1972ApJ...171..463S} {171, 463}

\bibitem[\protect\citeauthoryear{{Sullivan} et~al.,}{{Sullivan}
  et~al.}{2006}]{Sullivan06}
{Sullivan} M.,  et~al., 2006, \mn@doi [\apj] {10.1086/506137}, \href
  {https://ui.adsabs.harvard.edu/abs/2006ApJ...648..868S} {648, 868}

\bibitem[\protect\citeauthoryear{{Sullivan} et~al.,}{{Sullivan}
  et~al.}{2010}]{Sullivan2010}
{Sullivan} M.,  et~al., 2010, \mn@doi [\mnras]
  {10.1111/j.1365-2966.2010.16731.x}, \href
  {https://ui.adsabs.harvard.edu/abs/2010MNRAS.406..782S} {406, 782}

\bibitem[\protect\citeauthoryear{{Swann} et~al.,}{{Swann}
  et~al.}{2019}]{Tides2019}
{Swann} E.,  et~al., 2019, \mn@doi [The Messenger] {10.18727/0722-6691/5129},
  \href {https://ui.adsabs.harvard.edu/abs/2019Msngr.175...58S} {175, 58}

\bibitem[\protect\citeauthoryear{{The LSST Dark Energy Science Collaboration}
  et~al.,}{{The LSST Dark Energy Science Collaboration}
  et~al.}{2018}]{Mandelbaum18}
{The LSST Dark Energy Science Collaboration} et~al., 2018, arXiv e-prints,
  \href {https://ui.adsabs.harvard.edu/abs/2018arXiv180901669T} {p.
  arXiv:1809.01669}

\bibitem[\protect\citeauthoryear{{Tonry} et~al.,}{{Tonry}
  et~al.}{2003}]{Tonry03}
{Tonry} J.~L.,  et~al., 2003, \mn@doi [\apj] {10.1086/376865}, \href
  {https://ui.adsabs.harvard.edu/abs/2003ApJ...594....1T} {594, 1}

\bibitem[\protect\citeauthoryear{{Tripp}}{{Tripp}}{1998}]{Tripp98}
{Tripp} R.,  1998, \aap, \href
  {https://ui.adsabs.harvard.edu/abs/1998A&A...331..815T} {331, 815}

\bibitem[\protect\citeauthoryear{{Uddin}, {Mould}, {Lidman}, {Ruhlmann-Kleider}
   \& {Zhang}}{{Uddin} et~al.}{2017}]{Uddin17}
{Uddin} S.~A.,  {Mould} J.,  {Lidman} C.,  {Ruhlmann-Kleider} V.,   {Zhang}
  B.~R.,  2017, \mn@doi [\apj] {10.3847/1538-4357/aa8df7}, \href
  {https://ui.adsabs.harvard.edu/abs/2017ApJ...848...56U} {848, 56}

\bibitem[\protect\citeauthoryear{Vincenzi, Sullivan, Firth, Guti{\'{e} }rrez,
  Frohmaier, Smith, Angus  \& Nichol}{Vincenzi et~al.}{2019}]{Vincenzi19}
Vincenzi M.,  Sullivan M.,  Firth R.~E.,  Guti{\'{e} }rrez C.~P.,  Frohmaier
  C.,  Smith M.,  Angus C.,   Nichol R.~C.,  2019, \mn@doi [Monthly Notices of
  the Royal Astronomical Society] {10.1093/mnras/stz2448}, 489, 5802

\bibitem[\protect\citeauthoryear{{Walcher}, {Groves}, {Budav{\'a}ri}  \&
  {Dale}}{{Walcher} et~al.}{2011}]{Walcher11}
{Walcher} J.,  {Groves} B.,  {Budav{\'a}ri} T.,   {Dale} D.,  2011, \mn@doi
  [\apss] {10.1007/s10509-010-0458-z}, \href
  {https://ui.adsabs.harvard.edu/abs/2011Ap&SS.331....1W} {331, 1}

\bibitem[\protect\citeauthoryear{{Wiseman} et~al.,}{{Wiseman}
  et~al.}{2020}]{Wiseman20}
{Wiseman} P.,  et~al., 2020, \mn@doi [\mnras] {10.1093/mnras/staa2474}, \href
  {https://ui.adsabs.harvard.edu/abs/2020MNRAS.498.2575W} {498, 2575}

\bibitem[\protect\citeauthoryear{Wiseman, Sullivan, Smith  \& Popovic}{Wiseman
  et~al.}{2023}]{Wiseman23}
Wiseman P.,  Sullivan M.,  Smith M.,   Popovic B.,  2023, \mn@doi [Monthly
  Notices of the Royal Astronomical Society] {10.1093/mnras/stad488}, 520, 6214

\bibitem[\protect\citeauthoryear{Wolf et~al.,}{Wolf et~al.}{2016}]{Wolf16}
Wolf R.~C.,  et~al., 2016, \mn@doi [The Astrophysical Journal]
  {10.3847/0004-637x/821/2/115}, 821, 115

\bibitem[\protect\citeauthoryear{{Wood-Vasey} et~al.,}{{Wood-Vasey}
  et~al.}{2008}]{08Wood}
{Wood-Vasey} W.~M.,  et~al., 2008, \mn@doi [\apj] {10.1086/592374}, \href
  {https://ui.adsabs.harvard.edu/abs/2008ApJ...689..377W} {689, 377}

\bibitem[\protect\citeauthoryear{{Worthey}}{{Worthey}}{1994}]{Worthey94}
{Worthey} G.,  1994, \mn@doi [\apjs] {10.1086/192096}, \href
  {https://ui.adsabs.harvard.edu/abs/1994ApJS...95..107W} {95, 107}

\bibitem[\protect\citeauthoryear{{York} et~al.,}{{York} et~al.}{2000}]{York00}
{York} D.~G.,  et~al., 2000, \mn@doi [\aj] {10.1086/301513}, \href
  {https://ui.adsabs.harvard.edu/abs/2000AJ....120.1579Y} {120, 1579}

\bibitem[\protect\citeauthoryear{{de Jong} et~al.,}{{de Jong}
  et~al.}{2016}]{2016DeJong}
{de Jong} R.~S.,  et~al., 2016, in {Evans} C.~J.,  {Simard} L.,   {Takami} H.,
  eds,  Society of Photo-Optical Instrumentation Engineers (SPIE) Conference
  Series Vol. 9908, Ground-based and Airborne Instrumentation for Astronomy VI.
  p. 99081O, \mn@doi{10.1117/12.2232832}

\bibitem[\protect\citeauthoryear{{de Jong} et~al.,}{{de Jong}
  et~al.}{2019}]{deJong19a}
{de Jong} R.~S.,  et~al., 2019, \mn@doi [The Messenger]
  {10.18727/0722-6691/5117}, \href
  {https://ui.adsabs.harvard.edu/abs/2019Msngr.175....3D} {175, 3}

\makeatother
\end{thebibliography}

\newpage
\section*{Appendix} \label{Appendix}

\setlength{\extrarowheight}{3pt}
\begin{table*}
\centering
\begin{tabular}{|cc|ccc|cc|}
\hline
$r$-band & redshift & Template log(mass) & Phot log(mass) & Phot and spectroscopy & Phot with & Phot and spectroscopy\\
magnitude &  &(log[mass/$M_{\odot}$])  &  & log(mass) & Systematic error & With systematic error  \\  
\hline
18 & 0.1 & -     & $10.19_{10.11}^{10.28}$ & $10.25_{10.22}^{10.27}$ 
& $10.19_{10.09}^{10.30}$ & $10.25_{10.18}^{10.31}$\\
19 & 0.1 & -     & $9.79_{9.71}^{9.88}$    & $9.85_{9.82}^{9.87}$ 
& $9.79_{9.69}^{9.90}$ & $9.85_{9.78}^{9.91}$\\
20 & 0.1 & -      & $9.40_{9.32}^{9.49}$    & $9.45_{9.42}^{9.47}$    
& $9.40_{9.32}^{9.51}$ & $9.45_{9.38}^{9.51}$\\
21 & 0.1 & -      & $9.02_{8.92}^{9.14}$    & $9.05_{9.02}^{9.07}$    
& $9.02_{8.90}^{9.15}$ & $9.05_{8.98}^{9.11}$\\
22 & 0.1 & -      & $8.64_{8.53}^{8.79}$    & $8.65_{8.62}^{8.67}$
& $8.64_{8.51}^{8.80}$ & $8.65_{8.58}^{8.71}$\\
23 & 0.1 & -      & $8.25_{8.09}^{8.42}$    & $8.27_{8.21}^{8.43}$
& $8.25_{8.08}^{8.43}$ & $8.27_{8.19}^{8.44}$\\
24 & 0.1 & -      & $7.81_{7.52}^{8.05}$    & $7.91_{7.77}^{8.04}$
& $7.81_{7.51}^{8.06}$ & $7.91_{7.76}^{8.05}$\\
25 & 0.1 & -      & $7.32_{6.89}^{7.86}$    & $7.46_{7.18}^{7.68}$
& $7.32_{6.89}^{7.86}$ & $7.46_{7.17}^{7.69}$\\
\hline
18 & 0.3 & 11.46 & $11.43_{11.37}^{11.55}$ & $11.45_{11.42}^{11.51}$ 
& $11.43_{11.35}^{11.56}$ & $11.45_{11.38}^{11.53}$\\
19 & 0.3 & 11.05 & $11.04_{10.96}^{11.15}$ & $11.05_{11.02}^{11.11}$ 
& $11.04_{10.94}^{11.16}$ & $11.05_{10.98}^{11.13}$\\
20 & 0.3 & 10.65  & $10.65_{10.56}^{10.76}$ & $10.65_{10.62}^{10.71}$ 
& $10.65_{10.54}^{10.78}$ & $10.65_{10.58}^{10.72}$\\
21 & 0.3 & $\star\:10.25_{10.24}^{10.26}\:\star$ & $10.26_{10.15}^{10.37}$ & $10.25_{10.22}^{10.31}$ 
& $10.26_{10.13}^{10.39}$ & $10.25_{10.18}^{10.33}$ \\
22 & 0.3 & 9.85   & $9.88_{9.75}^{10.00}$   & $9.85_{9.82}^{9.91}$
& $9.88_{9.74}^{10.01}$ & $9.85_{9.78}^{9.93}$\\
23 & 0.3 & 9.44   & $9.52_{9.35}^{9.66}$    & $9.48_{9.40}^{9.63}$
& $9.52_{9.34}^{9.67}$ & $9.48_{9.38}^{9.64}$\\
24 & 0.3  & 9.04  & $9.13_{8.86}^{9.36}$    & $9.14_{8.98}^{9.28}$
& $9.13_{8.85}^{9.37}$ & $9.14_{8.97}^{9.29}$\\
25 & 0.3 & 8.64   & $8.68_{8.16}^{9.26}$    & $8.76_{8.52}^{8.99}$
& $8.68_{8.16}^{9.26}$ & $8.76_{8.51}^{9.00}$\\
\hline
18 & 0.5 & -     & $12.13_{12.05}^{12.38}$ & $12.20_{12.18}^{12.26}$ 
& $12.13_{12.03}^{12.39}$ & $12.20_{12.14}^{12.28}$\\
19 & 0.5 & -     & $11.75_{11.65}^{11.96}$ & $11.80_{11.78}^{11.86}$ 
& $11.75_{11.63}^{11.97}$ & $11.80_{11.74}^{11.88}$\\
20 & 0.5 & -      & $11.37_{11.26}^{11.57}$ & $11.40_{11.38}^{11.46}$ 
& $11.37_{11.24}^{11.58}$ & $11.40_{11.34}^{11.48}$\\
21 & 0.5 & -      & $10.99_{10.86}^{11.19}$ & $11.00_{10.98}^{11.06}$ 
& $10.99_{10.85}^{11.20}$ & $11.00_{10.94}^{11.08}$\\
22 & 0.5 & -      & $10.62_{10.47}^{10.81}$ & $10.60_{10.58}^{10.66}$
& $10.62_{10.46}^{10.82}$ & $10.60_{10.54}^{10.68}$\\
23 & 0.5 & -      & $10.28_{10.09}^{10.48}$ & $10.22_{10.16}^{10.37}$
& $10.28_{10.08}^{10.49}$ & $10.22_{10.14}^{10.38}$\\
24 & 0.5 & -      & $9.97_{9.70}^{10.25}$  & $9.88_{9.73}^{10.04}$
& $9.97_{9.69}^{10.26}$ & $9.88_{9.72}^{10.05}$\\
25 & 0.5 & -      & $9.59_{9.13}^{10.10}$   & $9.55_{9.33}^{9.79}$
& $9.59_{9.13}^{10.10}$ & $9.55_{9.32}^{9.80}$\\
\hline
18 & 0.7 & -     & $12.98_{12.70}^{13.06}$ & $12.88_{12.86}^{12.88}$ 
& $12.98_{12.69}^{13.08}$ & $12.88_{12.82}^{12.94}$\\
19 & 0.7 & -     & $12.58_{12.32}^{12.67}$ & $12.48_{12.46}^{12.48}$ 
& $12.58_{12.31}^{12.69}$ & $12.48_{12.42}^{12.54}$\\
20 & 0.7 & -      & $12.18_{11.96}^{12.30}$ & $12.08_{12.06}^{12.08}$ 
& $12.18_{11.95}^{12.31}$ & $12.08_{12.02}^{12.14}$\\
21 & 0.7 & -      & $11.79_{11.59}^{11.92}$ & $11.68_{11.66}^{11.68}$
& $11.79_{11.58}^{11.93}$ & $11.68_{11.62}^{11.74}$\\
22 & 0.7 & -      & $11.43_{11.23}^{11.58}$ & $11.28_{11.26}^{11.28}$
& $11.43_{11.22}^{11.59}$ & $11.28_{11.22}^{11.34}$\\
23 & 0.7 & -      & $11.05_{10.86}^{11.24}$ & $10.88_{10.85}^{11.03}$
& $11.05_{10.85}^{11.25}$ & $10.88_{10.81}^{11.04}$\\
24 & 0.7 & -      & $10.70_{10.45}^{10.96}$ & $10.54_{10.43}^{10.71}$
& $10.70_{10.44}^{10.97}$ & $10.54_{10.41}^{10.72}$\\
25 & 0.7 & -      & $10.36_{9.91}^{10.85}$  & $10.22_{10.01}^{10.45}$
& $10.36_{9.91}^{10.85}$ & $10.22_{10.00}^{10.46}$\\
\hline
\end{tabular}
\caption{The best-fit $\log (M_\ast)$ determined by FAST for an elliptical galaxy using photometry and 4MOST with photometry at various magnitudes and redshifts. The $\log M_\ast$ maximum is the 1-$\sigma$ upper limit on $M_\ast$ determined by FAST. Likewise, the log ($M_\ast$) minimum is the 1-$\sigma$ lower limit. The ranges reported are the 68th percentile ranges from FAST. A maximum uncertainty was introduced for 4MOST objects with a signal-to-noise that does not reach 3, to prevent the values from being unrealistically small. The first column shows our template values. The value at magnitude 21, z=0.3 (highlighted with stars) was calculated as described in section \ref{making template}. The remaining values do not have an uncertainty as the values were extrapolated from the calculated value at magnitude 21. The last two columns of the table show our results with a systematic error added in quadrature, based on the estimates reported by \citet{Pacifici22}.}
\label{Mass table el}
\end{table*}

\setlength{\extrarowheight}{3pt}
\begin{table*}
\centering
\begin{tabular}{|cc|ccc|cc|}
\hline
\label{Mass table sc}
$r$-band & redshift & Template log(mass) & Phot log(mass) & Phot and spectroscopy & Phot with & Phot and spectroscopy\\
magnitude &  &(log[mass/$M_{\odot}$])  &  & log(mass) & Systematic error & With systematic error  \\  
\hline
18 & 0.1 & -     & $10.12_{9.82}^{10.26}$ & $10.26_{10.22}^{10.26}$ 
& $10.12_{9.81}^{10.27}$ & $10.26_{10.19}^{10.32}$\\
19 & 0.1 & -     & $9.72_{9.42}^{9.86}$    & $9.86_{9.82}^{9.86}$ 
& $9.72_{9.41}^{9.87}$ & $9.86_{9.79}^{9.92}$\\
20 & 0.1 & -      & $9.32_{9.02}^{9.46}$    & $9.46_{9.42}^{9.46}$    
& $9.32_{9.01}^{9.47}$ & $9.46_{9.39}^{9.52}$\\
21 & 0.1 & -      & $8.90_{8.61}^{9.05}$    & $9.06_{9.02}^{9.06}$    
& $8.90_{8.60}^{9.06}$ & $9.06_{8.99}^{9.12}$\\
22 & 0.1 & -      & $8.47_{8.20}^{8.65}$    & $8.66_{8.62}^{8.66}$
& $8.47_{8.19}^{8.66}$ & $8.66_{8.59}^{8.72}$\\
23 & 0.1 & -      & $8.03_{7.76}^{8.23}$    & $8.24_{8.15}^{8.30}$
& $8.03_{7.75}^{8.24}$ & $8.24_{8.13}^{8.32}$\\
24 & 0.1 & -      & $7.54_{7.29}^{7.82}$    & $7.77_{7.58}^{7.89}$
& $7.54_{7.28}^{7.83}$ & $7.77_{7.57}^{7.90}$\\
25 & 0.1 & -      & $7.10_{6.69}^{7.60}$    & $7.23_{6.93}^{7.48}$
& $7.10_{6.67}^{7.60}$ & $7.23_{6.93}^{7.49}$\\
\hline
18 & 0.3 & 11.46 & $11.38_{11.26}^{11.46}$ & $11.41_{11.37}^{11.41}$ 
& $11.38_{11.25}^{11.48}$ & $11.41_{11.34}^{11.47}$\\
19 & 0.3 & 10.95 & $10.98_{10.86}^{11.06}$ & $11.01_{10.97}^{11.01}$ 
& $10.98_{10.85}^{11.08}$ & $11.01_{10.94}^{11.07}$\\
20 & 0.3 & 10.55  & $10.58_{10.45}^{10.66}$ & $10.61_{10.57}^{10.61}$ 
& $10.58_{10.44}^{10.68}$ & $10.61_{10.54}^{10.67}$\\
21 & 0.3 & $\star\:10.18_{10.13}^{10.19}\:\star$ & $10.17_{10.02}^{10.26}$ & $10.21_{10.17}^{10.21}$ 
& $10.17_{10.01}^{10.28}$ & $10.21_{10.14}^{10.27}$ \\
22 & 0.3 & 9.75   & $9.73_{9.49}^{9.86}$   & $9.81_{9.77}^{9.81}$
& $9.73_{9.48}^{9.87}$ & $9.81_{9.74}^{9.87}$\\
23 & 0.3 & 9.34   & $9.22_{9.00}^{9.42}$    & $9.41_{9.30}^{9.46}$
& $9.22_{8.99}^{9.43}$ & $9.41_{9.28}^{9.49}$\\
24 & 0.3  & 8.93  & $8.73_{8.47}^{9.00}$    & $8.93_{8.72}^{9.07}$
& $8.73_{8.46}^{9.01}$ & $8.93_{8.71}^{9.08}$\\
25 & 0.3 & 8.54  & $8.27_{7.80}^{8.81}$    & $8.39_{8.11}^{8.63}$
& $8.27_{7.80}^{8.81}$ & $8.39_{8.10}^{8.64}$\\
\hline
18 & 0.5 & -     & $12.07_{11.95}^{12.15}$ & $12.07_{12.04}^{12.07}$ 
& $12.07_{11.94}^{12.17}$ & $12.07_{12.00}^{12.13}$\\
19 & 0.5 & -     & $11.67_{11.55}^{11.75}$ & $11.67_{11.64}^{11.67}$ 
& $11.67_{11.54}^{11.77}$ & $11.67_{11.60}^{11.73}$\\
20 & 0.5 & -      & $11.27_{11.15}^{11.35}$ & $11.27_{11.24}^{11.27}$ 
& $11.27_{11.14}^{11.37}$ & $11.27_{11.20}^{11.33}$\\
21 & 0.5 & -      & $10.86_{10.74}^{10.95}$ & $10.87_{10.84}^{10.87}$ 
& $10.86_{10.73}^{10.97}$ & $10.87_{10.80}^{10.93}$\\
22 & 0.5 & -      & $10.44_{10.26}^{10.56}$ & $10.47_{10.44}^{10.47}$
& $10.44_{10.25}^{10.57}$ & $10.47_{10.40}^{10.53}$\\
23 & 0.5 & -      & $9.95_{9.71}^{10.13}$ & $10.07_{9.98}^{10.13}$
& $9.95_{9.70}^{10.14}$ & $10.07_{9.96}^{10.15}$\\
24 & 0.5 & -      & $9.44_{9.17}^{9.69}$  & $9.62_{9.43}^{9.76}$
& $9.44_{9.16}^{9.70}$ & $9.62_{9.43}^{9.77}$\\
25 & 0.5 & -      & $8.97_{8.44}^{9.44}$   & $9.08_{8.81}^{9.31}$
& $8.97_{8.44}^{9.44}$ & $9.08_{8.80}^{9.32}$\\
\hline
18 & 0.7 & -     & $12.60_{12.45}^{12.66}$ & $12.62_{12.61}^{12.62}$ 
& $12.60_{12.44}^{12.68}$ & $12.62_{12.56}^{12.68}$\\
19 & 0.7 & -     & $12.20_{12.05}^{12.26}$ & $12.22_{12.21}^{12.22}$ 
& $12.20_{12.03}^{12.28}$ & $12.22_{12.16}^{12.28}$\\
20 & 0.7 & -      & $11.80_{11.65}^{11.86}$ & $11.82_{11.81}^{11.82}$ 
& $11.80_{11.64}^{11.88}$ & $11.82_{11.76}^{11.88}$\\
21 & 0.7 & -      & $11.40_{11.25}^{11.46}$ & $11.42_{11.41}^{11.42}$
& $11.40_{11.24}^{11.48}$ & $11.42_{11.36}^{11.48}$\\
22 & 0.7 & -      & $11.00_{10.84}^{11.08}$ & $11.02_{11.01}^{11.02}$
& $11.00_{10.83}^{11.10}$ & $11.02_{10.96}^{11.08}$\\
23 & 0.7 & -      & $10.58_{10.36}^{11.70}$ & $10.61_{10.52}^{10.62}$
& $10.58_{10.35}^{10.71}$ & $10.61_{10.50}^{10.67}$\\
24 & 0.7 & -      & $10.10_{9.69}^{10.31}$ & $10.21_{10.05}^{10.29}$
& $10.10_{9.69}^{10.32}$ & $10.21_{10.03}^{10.31}$\\
25 & 0.7 & -      & $9.54_{8.90}^{9.94}$  & $9.70_{9.42}^{9.89}$
& $9.54_{8.90}^{9.94}$ & $9.70_{9.41}^{9.90}$\\
\hline
\end{tabular}
\caption{The same as Table \ref{Mass table el} but for an Sc galaxy.}
\end{table*}

\begin{figure*}
    \centering
    \includegraphics[scale=.29]{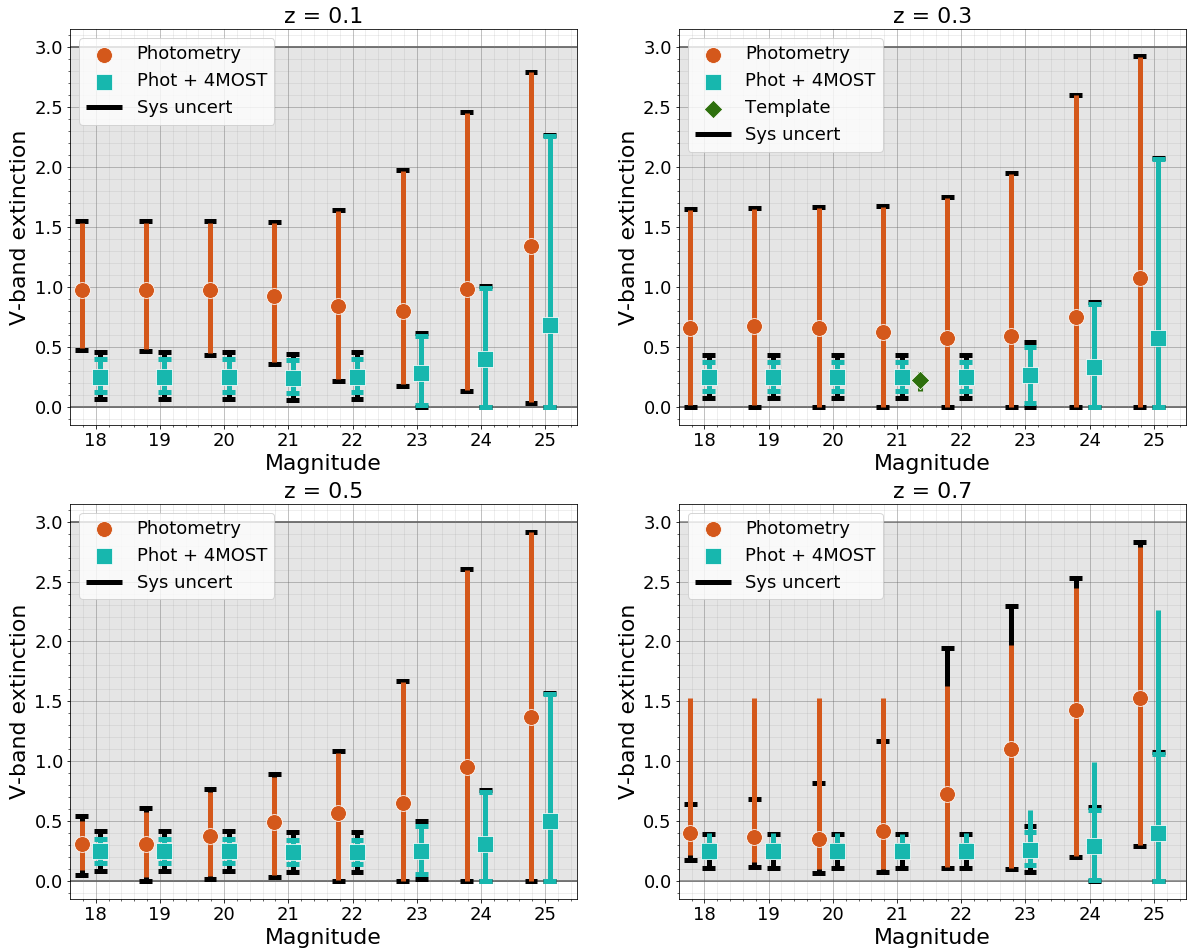}
    \caption{Simulated V-band extinction of an elliptical galaxy as a function of magnitude and redshift. The precision for this galaxy parameter also increases when 4MOST is used with photometry. The limits that V-band extinction value can be found between by FAST are shown by the grey shaded region. The limits are set at 0 magnitude and 3 magnitude.}
    \label{fig: 4panel av}
\end{figure*}

\begin{figure*}
    \centering
    \includegraphics[scale=.29]{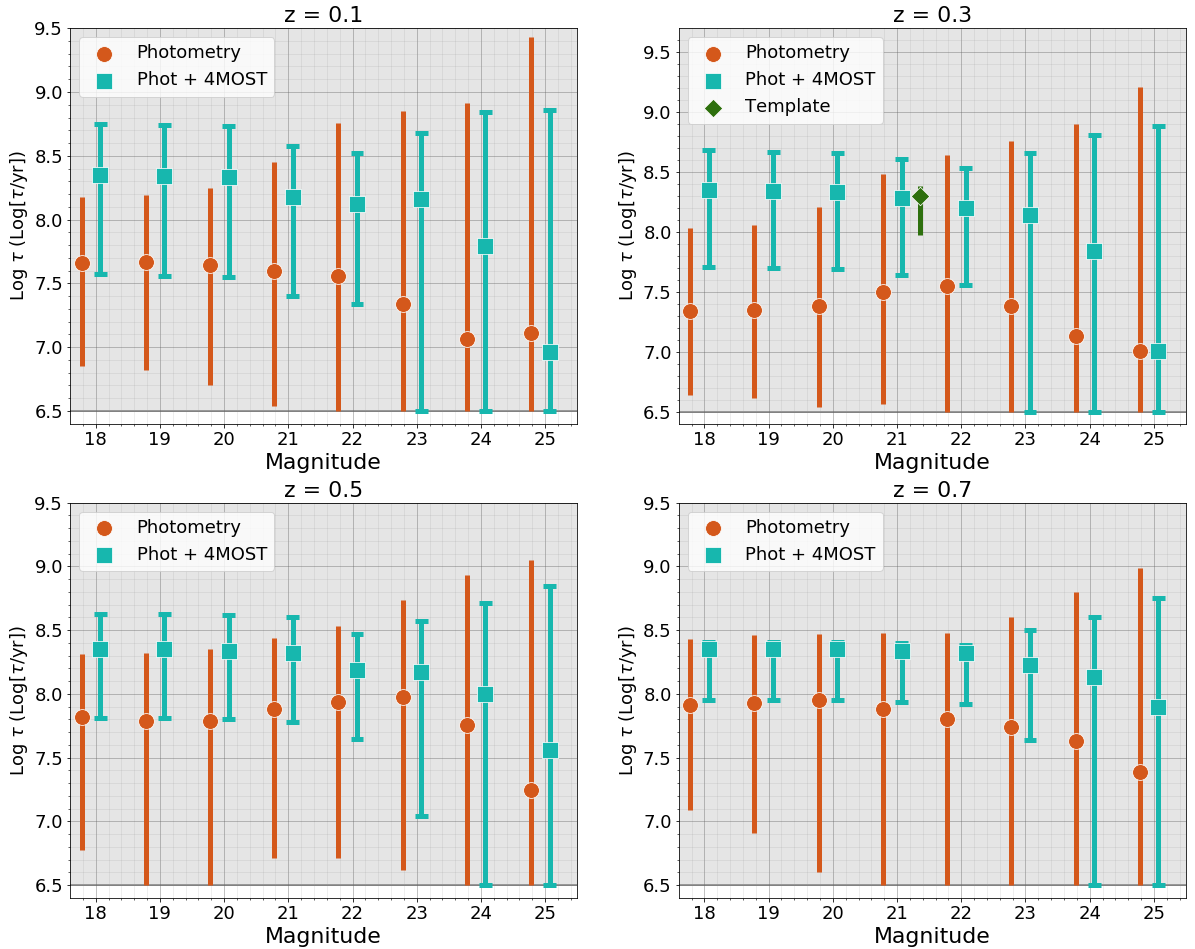}
    \caption{The simulated log (star formation timescale) results of an elliptical galaxy as a function of magnitude and redshift. The precision of the value is drastically improved at all simulated magnitudes and redshifts when 4MOST is used with photometry. However, at fainter magnitudes the $\tau$ value appears to be restricted by the lower limit. The limits of log $\tau$ covered in FAST's libraries is shown by the grey shaded region. The limits are set at 6.5 and 11 (log[$\tau$/yr]). }
    \label{fig: 4panel tau}
\end{figure*}

\begin{figure*}
    \centering
    \includegraphics[scale=.29]{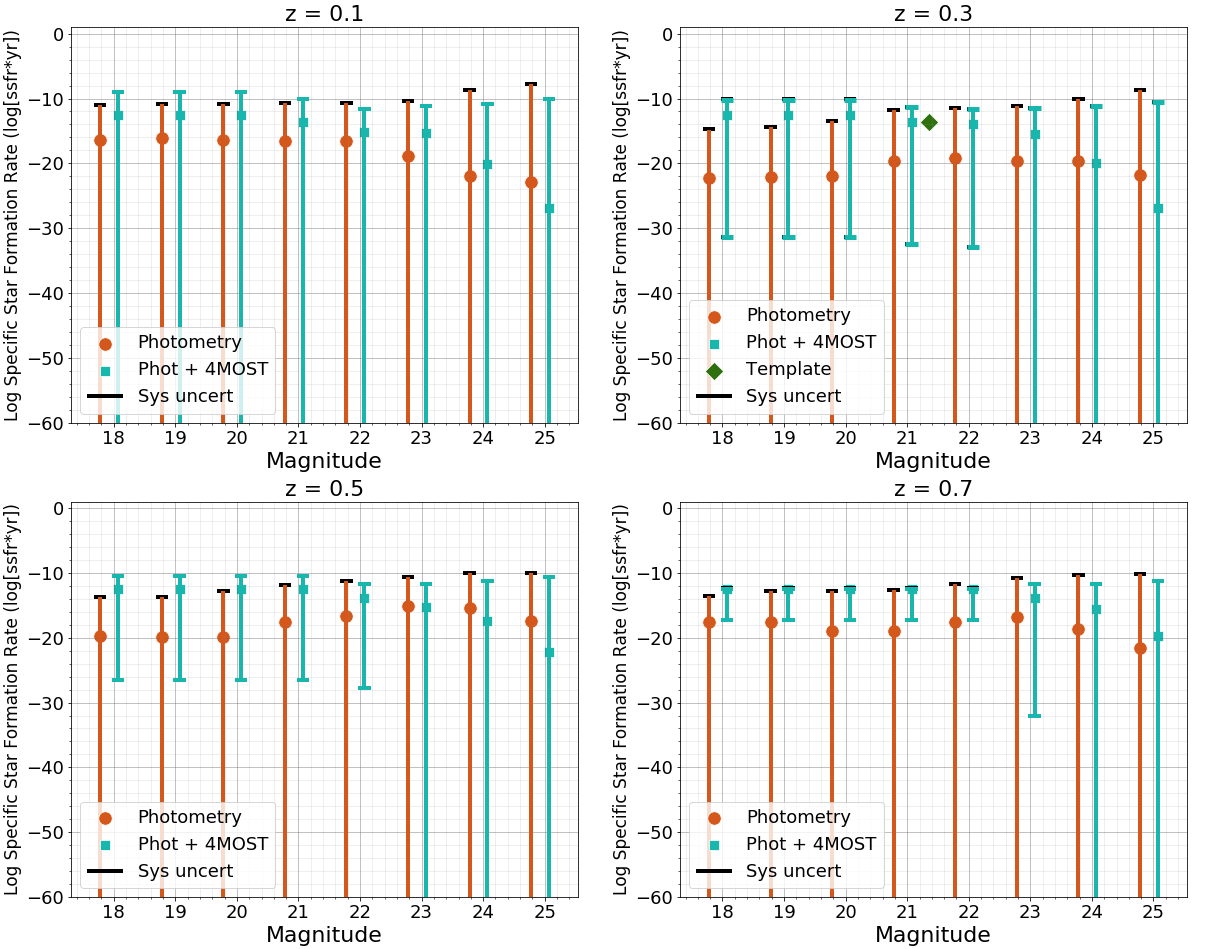}
    \caption{The simulated log Specific Star Formation Rate of an elliptical galaxy as a function of magnitude and redshift. The specific star formation rate sees an improvement at brighter magnitudes, for redshifts 0.3, 0.5 and 0.7. This is expected as it matches the host-galaxy mass and star formation rate results, which SSFR is a ratio of.}
    \label{fig: 4panel SSFR}
\end{figure*}

\begin{figure*}
    \centering
    \includegraphics[scale=.29]{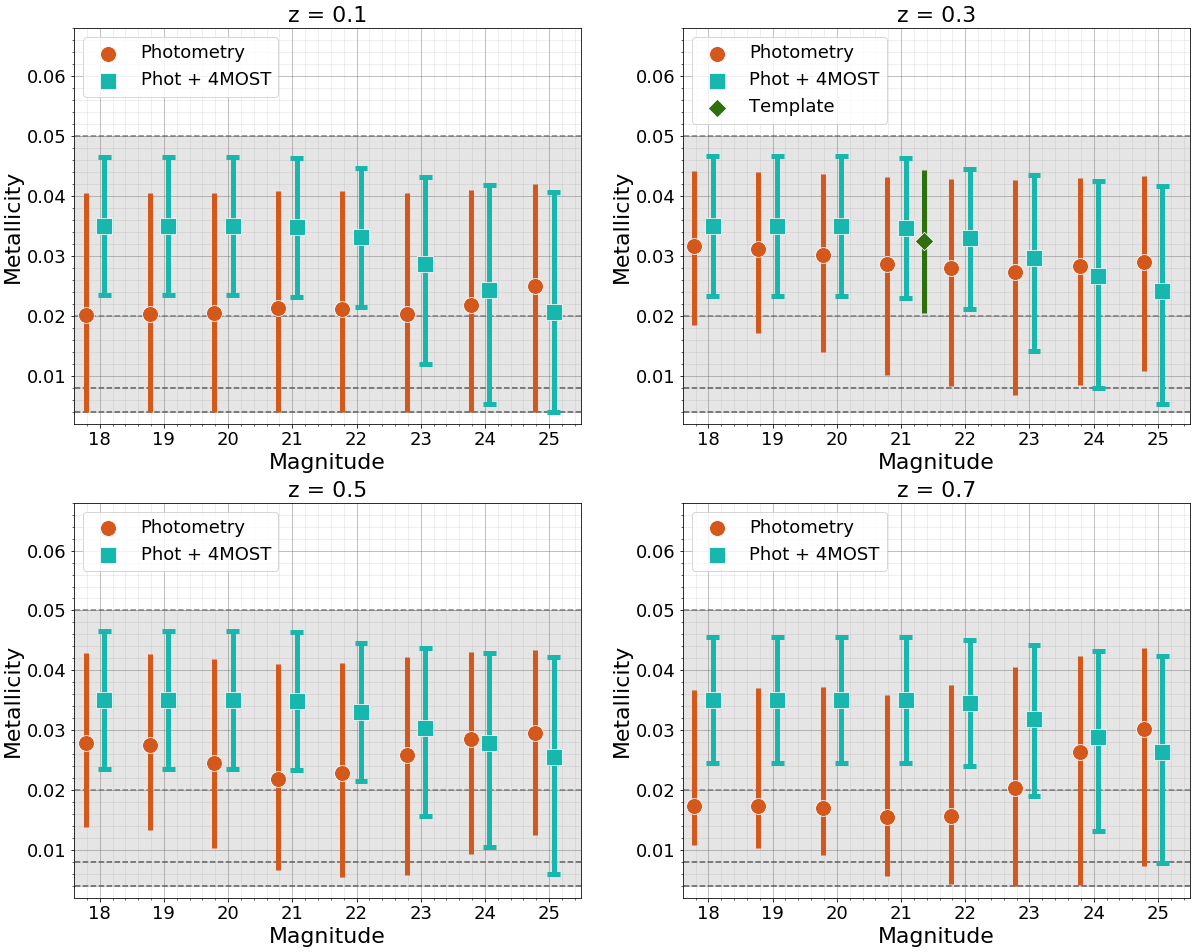}
    \caption{Simulated metallicity of of an elliptical galaxy as a function of magnitude and redshift. In the fit, metallicity was constrained to be one of four values, represented by the dashed lines. The precision slightly increases for most redshifts and magnitudes, but some of the fainter magnitudes have larger uncertainty ranges. The possible values were 0.004, 0.008, 0.02, 0.05.}
    \label{fig: 4panel metal}
\end{figure*}

\indent Additional galaxy parameters were investigated to observe the effect our method could have on the measured uncertainty. Uncertainties in V-band extinction see improvements at all simulated redshifts and magnitudes. The results for this can be seen in Figure \ref{fig: 4panel av}. The uncertainties measured on the V-band extinction value were reduced by 51 -- 95\% for magnitudes $\le$ 22 when using phot + 4MOST, whilst at fainter magnitudes there is a reduction to the uncertainties of 18 -- 87\%.  The star formation timescale ($\tau$) sees an improvement to its uncertainty at all simulated magnitudes and redshifts. For phot + 4MOST, the uncertainty on the measurement of star formation timescale was reduced by 11 -- 77\% for magnitudes $\le$ 22. Whilst at magnitudes fainter than 22 the uncertainty was reduced by 3 -- 59\%. We found that at magnitudes 24 and 25 the uncertainty range is constricted by the lower limit possible within FAST. The results for star formation rate timescale can be seen in Figure \ref{fig: 4panel tau}. We found that the $\tau$ parameter uncertainty range can hit the lower limit of $\tau$ values available within FAST's libraries. We found the $\tau$ uncertainty would hit the lower limit for magnitude 25 for all simulated redshifts. It also occurred with magnitude 24 for $z$=0.1, 0.3 and 0.5. The lower limit restriction could prevent the fit from reaching the true value of other galaxy parameters. However, as this only happens at the faintest magnitudes it is not a major concern to this study and could be investigated further in a study of deep fields.
\\\indent The measurement of specific star formation rate sees an improvement at brighter magnitudes, aside from redshift 0.1, but this is expected as it is a ratio of host-galaxy mass and star formation rate. The specific star formation rate results can be seen in Figure \ref{fig: 4panel SSFR}. Metallicity sees an improvement at most simulated magnitudes and redshifts when 4MOST spectroscopy is used. At magnitudes $\le$ 22 there is a reduction in uncertainty of 9 -- 37\%. However, at magnitudes 24 and 25 the uncertainties are the same size or even larger than when only photometry is used. Our template is a bright galaxy which has been magnitude normalised to fainter magnitudes. As we are calculating many parameters at the same time, it is possible that FAST does not cover the parameter space for this artificial combination of properties. The results for metallicity can be seen in Figure \ref{fig: 4panel metal}.

\bsp	
\label{lastpage}
\end{document}